\documentclass[letterpaper, 10 pt, conference]{ieeeconf}  % Comment this line out if you need a4paper

\IEEEoverridecommandlockouts                              
\usepackage{cite}
\usepackage{amsmath,amssymb,amsfonts}
\usepackage{algorithmic}
\usepackage[linesnumbered,ruled,vlined]{algorithm2e}
\usepackage{caption}
\usepackage{subcaption}
\usepackage{multirow, array}
\usepackage{graphicx}
\usepackage{textcomp}
\usepackage{xcolor}
\usepackage{booktabs}
\usepackage{makecell}
\usepackage{svg}

\usepackage{adjustbox}
\usepackage{pdflscape} % For landscape mode
\usepackage{booktabs}
\usepackage[normalem]{ulem}
\useunder{\uline}{\ul}{}
\usepackage{tabularx}
  {\list{}{\leftmargin=1em\rightmargin=1em}\item\relax\textbf{Key Finding:}\em}%
  {\endlist}

\title{\LARGE \bf
CSTS: A Canonical Security Telemetry Substrate \\
for AI-Native Cyber Detection
}

\author{Abdul Rahman\\
        arahman@alum.howard.edu
}

\begin{document}

\maketitle

\thispagestyle{empty}
\pagestyle{empty}

%%%%%%%%%%%%%%%%%%%%%%%%%%%%%%%%%%%%%%%%%%%%%%%%%%%%%%
\begin{abstract}
Cybersecurity data remains fragmented across vendors, formats, schemas, and deployment environments, forcing AI and analytics programs to spend disproportionate effort on ingestion, normalization, and brittle source-specific engineering. This paper introduces the Canonical Security Telemetry Substrate (CSTS), a canonical, AI-ready telemetry foundation designed to harmonize heterogeneous cyber data into a common representation over persistent entities, typed relations, events, temporal state, and provenance. CSTS is intended to move cybersecurity analytics beyond ad hoc record normalization toward a reusable substrate that supports anomaly detection, graph learning, forecasting, behavior-based modeling, and agentic cyber AI. We formalize the core design principles of CSTS, define its representational components, and explain how it preserves source-specific nuance through explicit mappings and extensible metadata while still enabling portable downstream inference. We further position CSTS as a cloud-agnostic and deployment-agnostic substrate suitable for on-prem, hybrid, and multi-cloud environments. The result is a unifying telemetry model that reduces the blue-collar burden of cyber data engineering and creates a clearer path to scalable, interoperable, and model-agnostic cyber AI.
\end{abstract}


%%%%%%%%%%%%%%%%%%%%%%%%%%%%%%%%%%%%%%%%%%%%%%%%%%%

\section{Introduction}
\label{sec:intro}

Machine learning, graph-based analytics, and representation-learning methods are now widely used in cybersecurity for tasks such as lateral movement detection, anomaly discovery, ransomware identification, and zero-day threat detection \cite{zdt_flow, zdt_metric, ransomware_alert_graph, lateral_movement_uba}. In controlled settings, these approaches can achieve strong performance; however, operational deployment remains difficult. Models that work well in one enterprise environment frequently degrade when moved to another, even when the underlying threat behaviors are similar. This degradation often persists across changes in topology, telemetry vendors, collection pipelines, and logging configurations, suggesting that the limiting factor is not model architecture alone.

A primary source of deployment friction is telemetry representation. Enterprise security data remains fragmented across heterogeneous vendor schemas, event-centric formats, and environment-specific pipelines. Normalization frameworks such as the Open Cybersecurity Schema Framework (OCSF) \cite{ocsf} reduce syntactic inconsistency and improve cross-vendor field alignment, but they remain fundamentally event-focused: higher-order constructs required for portable AI systems---persistent entity identity, typed relational semantics, explicit temporal state, and stable behavioral context---are typically reconstructed downstream on a per-environment basis. As a result, detection systems repeatedly re-implement identity stitching, relationship extraction, graph construction, and feature logic for each deployment, creating brittle coupling between detection logic and telemetry idiosyncrasies.

We view this as a \emph{representational gap}: the absence of a canonical abstraction layer between raw telemetry ingestion and downstream inference. Without such a substrate, portability failures conflate multiple mechanisms. Some failures arise from \emph{schema instability} (e.g., missing fields, renamed columns, vendor drift, parser mismatch). Others arise even when schema is nominally aligned: behavioral and graph-derived features can change their meaning under domain shift, producing \emph{semantic orientation instability}, in which class-conditional feature direction or interpretation reverses across environments. Distinguishing these mechanisms is essential for building robust AI-native security systems.

In this work, we introduce the \emph{Canonical Security Telemetry Substrate (CSTS)}, an entity-first, relationship-explicit representation layer that formalizes (i) persistent canonical identities, (ii) typed interactions as first-class relations, (iii) temporally indexed state, and (iv) provenance-preserving mappings from heterogeneous upstream sources. CSTS is designed as a stable integration boundary: heterogeneous telemetry sources are mapped into canonical entities, relations, events, and state updates via thin adapters, while downstream detection, graph, forecasting, and representation-learning systems consume CSTS primitives rather than vendor-specific event fields. In this sense, CSTS is not merely a normalization schema; it is an AI-ready substrate intended to reduce the repeated reinvention of telemetry semantics across environments and use cases.

A key design goal of CSTS is portability. The substrate is intended to function across cloud, on-prem, and hybrid environments, and across heterogeneous data producers, by enforcing a stable representational core while preserving source-specific nuance through governed mappings and extensible metadata. This design allows ingestion logic to evolve independently from downstream analytics and creates a clearer path toward plug-and-play model deployment: once telemetry is mapped into CSTS, downstream AI systems can operate over canonical objects rather than source-specific record layouts.

We empirically evaluate CSTS under both controlled and external settings. In a synthetic two-environment benchmark (Env~A$\rightarrow$Env~B) designed to isolate representational effects, CSTS materially reduces transfer degradation for identity-centric detection (lateral movement) relative to an event-centric baseline and remains operable under targeted schema perturbations that cause the baseline pipeline to fail. For flow-centric zero-day detection, CSTS preserves schema alignment but reveals a distinct portability boundary: directional stability of graph-derived statistics under domain shift remains a separate modeling challenge. Finally, on public real-log corpora we include external robustness/smoke tests and a producer-divergence case study (TC~E3) to validate pipeline behavior under realistic telemetry heterogeneity and to surface cross-producer distributional mismatch under leakage-safe thresholding.

This paper makes three contributions:
\begin{enumerate}
    \item \textbf{Canonical substrate.} We formalize CSTS as a governed, entity-relational telemetry layer with explicit identity persistence, typed relationship semantics, temporal indexing, and provenance-aware canonicalization.
    \item \textbf{Portability decomposition.} We articulate a portability framework that separates \emph{schema stability} (representation and ingestion invariants) from \emph{semantic orientation stability} (feature-direction and behavioral-meaning invariance under domain shift).
    \item \textbf{Empirical evidence.} We provide controlled transfer and schema-perturbation experiments demonstrating improved portability for identity-centric detection, and we document a producer-divergence artifact illustrating that schema alignment alone does not guarantee cross-producer calibration or semantic alignment.
\end{enumerate}
%%%%%%%%%%%%%%%%%%%%%%%%%%%%%%%%%%%%%%%%%%%%%%
\section{Background}

Modern cybersecurity architectures aggregate telemetry from heterogeneous sources including network infrastructure, identity providers, endpoint agents, cloud control planes, SaaS platforms, and threat intelligence feeds. Although advances in normalization frameworks and machine learning--based detection have improved large-scale analysis, persistent deployment friction remains. Integration overhead, representational instability, and cross-environment degradation continue to limit reliable AI adoption in production environments.

We review three relevant foundations: (i) event-centric normalization frameworks, (ii) machine learning--based detection systems, and (iii) empirical deployment gaps observed in practice.

\subsection{Event-Centric Normalization}

Enterprise telemetry is fragmented across vendors, schemas, and collection pipelines. To address syntactic inconsistency, normalization frameworks such as the Open Cybersecurity Schema Framework (OCSF) \cite{ocsf}, STIX/TAXII, Elastic Common Schema (ECS), and related standards align field names and event structures across products.

These frameworks improve interoperability by standardizing event-level attributes such as timestamps, source/destination identifiers, event categories, and metadata fields. In practice, normalization reduces parsing complexity and enables cross-tool ingestion into SIEM, observability, and data lake environments.

However, event-centric normalization primarily solves \emph{schema alignment}. It does not provide a stable higher-order semantic abstraction for AI-native systems. Persistent entity identity, typed relationships among entities, temporal behavioral state, longitudinal role modeling, and cross-source provenance are typically reconstructed downstream inside detection pipelines. Consequently, identity stitching is repeatedly reimplemented for each deployment, graph-construction logic becomes bespoke and tightly coupled to specific detectors, and behavioral aggregation varies across environments. This fragmentation increases operational overhead and introduces instability into downstream learning systems.

Thus, while normalization frameworks reduce syntactic heterogeneity, they do not define a canonical entity-relational substrate suitable for stable AI, graph, and representation-learning workflows across heterogeneous deployments.

\subsection{Machine Learning--Based Detection}

Machine learning methods are widely applied to cybersecurity tasks including anomaly detection, lateral movement identification, zero-day discovery, ransomware detection, and insider threat monitoring. Supervised classifiers, unsupervised anomaly detectors, graph neural networks, autoencoders, and metric-learning approaches have demonstrated strong performance in controlled settings.

Graph-based models capture relational structure among users, hosts, processes, files, and network flows. Autoencoder architectures model baseline behavior and identify deviations indicative of novel attacks. Metric-learning frameworks structure latent representations to improve separation of attack classes and support transfer across threat categories.

Despite their architectural sophistication, these systems often rely on environment-specific preprocessing, feature engineering, and identity-resolution rules. Relationship semantics and behavioral features are frequently tightly coupled to telemetry idiosyncrasies. As a result, feature definitions become unstable across deployments, transfer performance degrades under topology or vendor changes, and model portability often requires retraining, manual adapter work, or repeated feature redesign. These effects collectively undermine the reliability of AI-based detection systems in heterogeneous enterprise environments. The lesson is not that model architecture is irrelevant, but that architectural advances alone are insufficient without a stable representational boundary between ingestion and inference.

\subsection{Deployment Gaps}

Operational deployment reveals recurring friction beyond model accuracy. Integrating new telemetry sources requires adapter development, identity reconciliation, relation extraction, and reconstruction of higher-order behavioral abstractions. Vendor updates, parser changes, and schema drift introduce ongoing maintenance burden.

Furthermore, graph structures used by detection systems are often rebuilt independently within each pipeline rather than derived from a shared canonical representation. This increases operational complexity and creates tight coupling between ingestion logic and modeling logic. It also impedes reproducibility: the same nominal detection idea may be implemented differently across environments because the underlying telemetry abstractions are reconstructed ad hoc each time.

Cross-environment generalization remains particularly challenging. Models trained in one enterprise topology may degrade when deployed in another due to differences in identity semantics, role distributions, telemetry coverage, and behavioral context---even when schema fields are nominally aligned. This suggests that portability failures arise not only from missing or mismatched fields, but from deeper semantic instability in how behavior is represented.

Collectively, these gaps indicate the absence of a canonical telemetry substrate that formalizes persistent identity, typed relationships, temporal state, and provenance independently of vendor-specific event formats. CSTS is proposed as that missing layer.

%%%%%%%%%%%%%%%%%%%%%%%%%%%%%%%%%%%%%%%%%%%%%%%%%%%%%%%%%%%%%%%%%%%%%%%%%%%%%%%%
\section{Related Work}

AI-driven cybersecurity systems sit at the intersection of enterprise telemetry architecture, schema normalization, graph-based reasoning, and learning-based detection. Substantial progress has been made across these areas. However, most prior work either assumes the availability of stable, semantically coherent telemetry representations or treats representational issues as downstream engineering details rather than as a first-class research problem. We synthesize the relevant literature and identify a persistent representational gap: the absence of a vendor-agnostic, entity-first, relationship-explicit substrate designed explicitly to support portable AI deployment across heterogeneous enterprise environments.

\subsection{Telemetry Security Architecture and Enterprise Deployment Context}

Enterprise telemetry infrastructures are increasingly recognized as high-value assets requiring layered security controls. Kalibjian emphasizes that telemetry post-processing environments contain sensitive and strategically valuable data, arguing for encryption, key management, and compliance-aligned protections for telemetry at rest and in motion~\cite{kalibjian2017telemetrysecurity}. Zegeye and Odejobi propose a cybersecurity architecture for telemetry networks modeled after ICS/SCADA systems, emphasizing segmentation, zone-based control, and boundary management across enterprise and field layers~\cite{zegeye2022telemetryarch}. Dean \emph{et al.} further extend this perspective with a networked telemetry security architecture integrating layered controls across distributed enterprise domains and positioning telemetry infrastructure as a foundation for advanced analytics and AI-driven monitoring~\cite{dean2024networkedtelemetry}.

These works contribute essential operational realism. They clarify how telemetry is generated, transmitted, protected, and governed in real environments. However, telemetry is treated primarily as protected data-in-motion or secured operational infrastructure rather than as a formally defined representational layer for AI systems. Canonical identity persistence, typed inter-entity semantics, provenance-preserving relation construction, and substrate-level temporal abstraction are not explicit design objects in this literature. As a result, secure telemetry architecture is addressed, but AI-ready telemetry representation remains largely implicit.

\subsection{Telemetry Normalization and Observability Frameworks}

Normalization frameworks such as the Open Cybersecurity Schema Framework (OCSF)~\cite{ocsf}, Elastic Common Schema (ECS)~\cite{ecs}, and OpenTelemetry~\cite{opentelemetry} have significantly improved cross-vendor interoperability through standardized event schemas and attribute alignment. These frameworks reduce ingestion friction, simplify SIEM and observability integration, and enable more consistent field naming across heterogeneous sources.

Rongali demonstrates that OpenTelemetry instrumentation can be extended with enriched spans and security-oriented attributes to enhance detection fidelity~\cite{rongali2025opentelemetry}. Such efforts show that structured instrumentation improves contextual signal quality and can move observability data closer to security analytics use cases.

However, these frameworks remain fundamentally event-centric. They standardize event syntax and attribute organization, but they do not formalize persistent entity identity across sessions and vendors, canonical inter-entity relationships, temporal behavioral state, or governed provenance-aware mappings suitable for portable AI inference. Higher-order abstractions required for graph learning, behavioral modeling, and cross-environment transfer are therefore reconstructed downstream inside environment-specific pipelines. Normalization reduces syntactic variance, but it does not by itself guarantee semantic stability.

\subsection{Telemetry Representation and Scalable Security Analytics}

A smaller body of work has attempted to elevate telemetry beyond raw event streams by introducing more structured or relational representations. Taylor \emph{et al.} introduce SysFlow, a compact telemetry format that lifts low-level system events into a flow-centric, object-relational abstraction linking processes, files, and network endpoints~\cite{taylor2020sysflow}. SysFlow improves interpretability and reduces storage overhead by compressing event granularity into relational summaries, and it points in an important direction: telemetry becomes more useful for analytics when relationships and object context are represented explicitly.

While aligned with the broader goal of relational abstraction, SysFlow is primarily host-centric and does not by itself generalize across heterogeneous enterprise domains such as identity providers, SaaS platforms, cloud control planes, email systems, or threat intelligence sources. Nor does it define governance mechanisms for cross-domain identity persistence, canonical relationship ontologies, or deployment-agnostic representational contracts suitable for stable AI deployment. In that sense, representation compression improves efficiency and local semantics, but it does not yet establish a universal substrate for cross-environment portability.

\subsection{Learning-Based Detection and Graph-Oriented Security Systems}

Machine learning is central to modern intrusion detection. Okonkwo surveys advanced intrusion detection approaches incorporating behavioral modeling and AI-enhanced anomaly detection in enterprise telemetry networks~\cite{okonkwo2024advancedids}. Zero-day detection research emphasizes anomaly modeling and behavioral generalization for detecting previously unseen attack patterns~\cite{zdt_flow,zdt_metric}. Metric-learning approaches structure latent representations to improve discrimination of malicious behavior and robustness of downstream decision boundaries~\cite{zdt_metric,martinez2024metricfused,koukoulis2025selfsup}.

Graph-based intrusion detection systems and knowledge-graph security models encode entities and interactions to enable contextual reasoning, attack-path reconstruction, and multi-stage campaign analysis~\cite{sarhan2021opencykg,sikos2023cyberkg,gilliard2024kgreasoning,zhou2024knowgraph}. These systems demonstrate that relational structure is essential for detecting lateral movement, privilege propagation, coordinated activity, and higher-order adversarial behavior that is poorly captured by isolated event statistics.

However, most graph and learning-based systems derive entity graphs, behavioral contexts, and feature abstractions from event streams in environment-specific ways. Identity resolution, relationship extraction, temporal slicing, and graph construction remain tightly coupled to vendor schemas and telemetry peculiarities. As a result, portability across enterprises is limited, and deployment frequently requires retraining, re-engineering, or environment-specific reconstruction of the representational layer. The limiting factor increasingly lies not in model capacity alone, but in representational instability.

\subsection{Telemetry Collection and Dataset Generation}

Repeatable dataset generation is also an important part of the ecosystem. Holeman \emph{et al.} introduce SETC, a containerized framework for repeatable exploit execution and telemetry collection to support controlled dataset generation and defensive model evaluation~\cite{holeman2024setc}. Such frameworks improve reproducibility, enable better experiment control, and make it easier to compare defensive systems under known attack scenarios.

However, improved dataset generation does not itself resolve representational heterogeneity in operational settings. Telemetry collected in controlled environments still requires identity reconciliation, relation extraction, temporal normalization, and environment-aware semantic abstraction before it can support portable deployment across heterogeneous enterprises. Reproducible capture solves part of the evaluation problem, but it does not solve the substrate problem.

\subsection{Feature Drift, Domain Adaptation, and Deployment Fragility}

Feature and concept drift are well-documented challenges in operational intrusion detection, where evolving traffic patterns, topology changes, and attack behavior shifts degrade model performance over time~\cite{shyaa2024driftids}. Broader surveys of distribution shift emphasize that deployed machine learning systems require continuous monitoring and adaptation in non-stationary environments~\cite{hinder2024driftsurveyA}.

In enterprise cybersecurity, distribution shift is compounded by schema variation, vendor heterogeneity, telemetry incompleteness, and inconsistent identity semantics across deployments. Existing mitigation strategies focus primarily on retraining, adaptive ensembles, or robustness techniques, but they rarely address representational stability at the telemetry substrate level. This is especially important because some deployment failures reflect missing or drifting fields, while others reflect shifts in the behavioral meaning of derived features even when schemas are nominally aligned.

This recurring deployment fragility suggests that a major bottleneck in AI-driven cybersecurity is architectural rather than purely algorithmic. Without a stable canonical substrate, portability failures at the representational layer are repeatedly pushed downstream into model retraining and feature repair.

\subsection{Synthesis: The Representational Gap}

Across telemetry security architecture, normalization frameworks, scalable representation formats, graph-based detection systems, and machine-learning deployment research, three themes consistently emerge:

\begin{enumerate}
    \item Enterprise telemetry is operationally complex and strategically sensitive, requiring secure, governed, and deployment-aware infrastructure~\cite{kalibjian2017telemetrysecurity,zegeye2022telemetryarch,dean2024networkedtelemetry}.
    \item Advanced detection methods rely on stable relational and behavioral abstractions, including persistent entity identity, typed inter-entity semantics, and cross-domain contextual modeling~\cite{zdt_metric,martinez2024metricfused,koukoulis2025selfsup,sarhan2021opencykg,sikos2023cyberkg}.
    \item Event-centric normalization frameworks improve ingestion and interoperability, but they do not formalize a canonical entity-relational substrate with governed identity persistence, temporal continuity, provenance-aware mappings, and stable representational contracts for AI portability~\cite{ocsf,ecs,opentelemetry,taylor2020sysflow,rongali2025opentelemetry}.
\end{enumerate}

No surveyed work defines a cross-domain, entity-first, relationship-explicit telemetry layer with persistent identity resolution, typed relational semantics, temporal state continuity, provenance preservation, and representational stability contracts designed explicitly for portable AI deployment across heterogeneous enterprise environments.

This representational gap motivates the Canonical Security Telemetry Substrate (CSTS), introduced next. CSTS is intended not merely as another normalization scheme, but as a canonical AI-ready substrate that reduces the repeated reinvention of telemetry semantics and creates a clearer path from ingestion to portable inference.
%%%%%%%%%%%%%%%%%%%%%%%%%%%%%%%%%%%%%%%%%%%%%%%%%%%%%%%%%%%%%%%%%%%%%%%%%%%%%%%%
\section{Design Principles of CSTS}
\label{sec:design_principles}

The Canonical Security Telemetry Substrate (CSTS) is designed as an AI-native representation layer that formalizes how heterogeneous enterprise security telemetry is structured before downstream modeling. Unlike event-centric normalization frameworks such as OCSF~\cite{ocsf}, ECS~\cite{ecs}, and OpenTelemetry~\cite{opentelemetry}, CSTS is not limited to field alignment. Its purpose is to define a stable, governed, and portable representational boundary over which AI systems can be built. In particular, CSTS is intended to convert fragmented source-specific telemetry into a canonical substrate over persistent entities, typed relationships, events, temporal state, and provenance, so that downstream models operate on consistent semantics rather than brittle vendor-specific event layouts.

A central design goal of CSTS is to reduce the repeated blue-collar burden of cybersecurity data engineering. In current practice, identity stitching, relation extraction, graph construction, behavioral aggregation, and feature stabilization are repeatedly reimplemented inside each downstream pipeline. CSTS shifts this burden upward into a shared substrate. To do so reliably, the substrate must satisfy not only semantic requirements but also deployment requirements: it must work across cloud, on-prem, and hybrid environments; it must preserve source-specific nuance without breaking downstream interfaces; and it must support governed evolution as vendors, pipelines, and use cases change. These requirements motivate the following design principles.

CSTS is guided by six design principles.

\subsection{Principle 1: Persistent Entity Identity}

CSTS adopts an entity-first paradigm: security-relevant objects are first-class substrate constructs rather than implicit artifacts reconstructed from events. Canonical entity types include users, hosts, processes, files, network endpoints and flows, credentials, cloud resources, binaries, services, devices, and externally sourced indicators. Each entity is assigned a persistent canonical identifier intended to remain stable across telemetry sources and over time.

This principle targets a central deployment failure mode in event-centric systems: identity instability. In heterogeneous environments, a single logical entity may appear under multiple identifiers across vendors and time---for example hostname aliases, principal-name variants, ephemeral cloud IDs, IP reassignment, or vendor-specific object naming. When identity is not stabilized, downstream pipelines repeatedly restitch entities, producing inconsistent graph construction, brittle features, and degraded cross-environment transfer. CSTS therefore makes identity persistence a substrate responsibility so that detection logic can remain vendor-agnostic and topology-portable.

Persistent identity is not equivalent to naive field deduplication. It requires governed identity resolution policies, explicit provenance, and the ability to maintain canonical identity even when upstream telemetry is incomplete or partially conflicting. In CSTS, identity resolution is thus treated as a core representational obligation rather than a convenience feature.

\subsection{Principle 2: Explicit Relationship Semantics}

Security-relevant behavior is fundamentally relational. CSTS therefore materializes typed relationships as explicit substrate constructs rather than deriving them ad hoc inside model pipelines. Relationship types are semantically constrained (e.g., \texttt{AUTHENTICATES\_TO}, \texttt{EXECUTES}, \texttt{SPAWNS}, \texttt{CONNECTS\_TO}, \texttt{WRITES}, \texttt{PRIV\_ESCALATES}) and define permissible source and target entity classes.

By enforcing relationship semantics at the substrate layer, CSTS yields a canonical entity-relational graph that is stable across deployments. Downstream systems consume pre-materialized interaction structure rather than rebuilding topology from raw logs, improving reproducibility and reducing environment-specific feature re-engineering. This also enables graph learning, knowledge-based reasoning, anomaly detection, and behavioral modeling methods to operate over consistent semantics without bespoke reconstruction logic.

Equally important, explicit relationship semantics make the substrate compositional. Once entities and relations are canonicalized, multiple downstream views can be generated from the same core representation: graph slices, sequence slices, macrostate windows, behavior summaries, and model-specific feature tables. This is one of the main reasons CSTS is designed as a substrate rather than as a narrow schema.

\subsection{Principle 3: Temporal Continuity and Behavioral State}

Enterprise security telemetry is inherently temporal: many detection paradigms depend on deviations from historical baselines, ordered sequences of actions, persistent role evolution, and event-to-state transitions over time. Event normalization records timestamped events, but typically leaves temporal joins, state reconstruction, and longitudinal behavioral modeling to downstream pipelines.

CSTS incorporates temporal continuity at the substrate boundary by representing events, relations, and state as time-indexed objects with explicit temporal validity. This supports rolling aggregation windows, behavioral baselining, sequence modeling, and trajectory-based reasoning in a uniform way across environments. By centralizing temporal semantics, CSTS reduces duplication of time-aggregation logic across detection systems and improves cross-topology comparability of behavioral features.

This principle also matters for future AI layers beyond conventional anomaly detection. Forecasting, sequence modeling, regime analysis, behavioral macrostate modeling, and agentic cyber reasoning all depend on stable temporal semantics. CSTS is designed so that these tasks can be built over a common time-aware substrate rather than over source-specific event stitching.

\subsection{Principle 4: Interface Stability Through Explicit Mappings}

A core engineering doctrine of CSTS is: \emph{make the interfaces identical and the reference mappings explicit}. Downstream AI systems should consume stable canonical contracts, while source-specific nuance should be captured in governed mapping profiles rather than leaking into every model and feature pipeline.

This principle means that vendor feeds, benchmark datasets, cloud-native logs, and future telemetry producers are adapted into the same CSTS objects through explicit, versioned mapping specifications. These mappings define how source fields become canonical entities, relationships, events, state updates, and provenance records. As a result, source-specific complexity is not ignored, but it is contained.

The practical consequence is deployment portability. An enterprise may run on AWS, Azure, GCP, on-prem, or hybrid infrastructure; it may use Palo Alto, CrowdStrike, Microsoft, Okta, Zeek, NetFlow, or custom producers; but once those sources are mapped into CSTS, downstream systems see the same canonical interfaces. This separation between source mappings and canonical interfaces is essential for scalable AI deployment and is one of the main distinctions between CSTS and ordinary field-normalization schemes.

\subsection{Principle 5: Governed Evolution, Feature Stability, and Extensibility}

A recurring cause of enterprise ML fragility is feature instability induced by schema drift, vendor updates, changing telemetry coverage, and evolving operational context. When features depend directly on raw source fields, even modest upstream changes can invalidate models or require costly reintegration.

CSTS addresses this through schema governance and feature stability contracts. Canonical behavioral categories---such as frequency, diversity, rarity/novelty, privilege spread, role deviation, interaction concentration, and temporal persistence---are defined over canonical entities, relations, and state rather than over vendor-specific raw fields. These abstractions are versioned independently of raw source schemas. Changes in upstream telemetry are therefore absorbed by adapter-layer mappings and canonicalization policies rather than by modifying downstream model interfaces.

At the same time, CSTS must remain extensible. The substrate is therefore designed with a strict core and extensible edges. The core defines required canonical semantics; extension zones preserve source-specific residuals, custom attributes, and future telemetry types without breaking the downstream interface. This allows CSTS to evolve without forcing continuous redesign of detection logic.

\subsection{Principle 6: Deployment Decoupling and AI-Ready Consumption}

CSTS decouples telemetry integration from detection and inference logic. Vendor feeds are mapped through thin adapters into canonical entities, relationships, events, state updates, and provenance; downstream consumers then operate on the uniform substrate. This reduces onboarding friction, confines environment-specific work to the mapping layer, and enables incremental adoption without requiring a full-stack redesign for every new telemetry source.

This principle is especially important for cloud and deployment portability. CSTS is intended to function across AWS, Azure, GCP, on-prem, and hybrid environments by keeping the canonical contracts fixed even when the underlying storage, orchestration, or compute stack changes. In this sense, CSTS is deployment-agnostic by design: cloud-specific ingestion differs at the edge, but the AI-facing substrate remains the same.

CSTS is also intentionally model-agnostic while remaining AI-ready. The substrate is designed to support supervised learning, unsupervised anomaly detection, graph learning, metric learning, forecasting, embedding-based retrieval, behavioral macrostate modeling, and future agentic workflows over shared canonical state. The objective is not merely to normalize telemetry, but to create a stable inference boundary over which many AI systems can be built and reused.

\subsection{Summary of Design Principles}

Taken together, these principles shift the representational burden from individual detection pipelines to a shared substrate. Persistent identity, explicit relationship semantics, temporal continuity, interface stability through explicit mappings, governed evolution, and deployment decoupling establish CSTS as a canonical telemetry substrate for portable AI-driven cybersecurity.

More concretely, CSTS is intended to do three things simultaneously: (i) harmonize heterogeneous telemetry into a common entity-relational form, (ii) preserve enough source-specific nuance to remain operationally faithful, and (iii) expose stable canonical interfaces that support downstream inference across cloud, on-prem, and hybrid environments. This is the sense in which CSTS is not just a schema proposal, but a governed AI-ready substrate designed to support scalable, interoperable, and model-agnostic cyber analytics.

%%%%%%%%%%%%%%%%%%%%%%%%%%%%%%%%%%%%%%%%%%%%%%%%%%%%%%%%%%%%%%%%%%%%%%%%%
\section{Formal Definition of CSTS}
\label{sec:formal_definition}

The Canonical Security Telemetry Substrate (CSTS) defines a stable, entity-centric abstraction over heterogeneous enterprise telemetry. Rather than modeling raw vendor events directly, CSTS models security data as a time-indexed attributed multigraph together with governed state and provenance layers:
\begin{equation}
G_t = (V_t, E_t).
\end{equation}
Here $V_t$ denotes the set of canonical entities valid at time $t$, and $E_t$ denotes the set of typed relationships among those entities. This abstraction shifts cybersecurity analytics from event-centric processing to entity--relationship reasoning while enforcing identity, structural, temporal, and governance invariants intended to improve portability across environments.

The core purpose of this section is to make explicit what CSTS is, what it guarantees, and what it does not guarantee. CSTS is not a replacement for all source-specific semantics, nor is it itself a model. It is a governed representational boundary that sits between heterogeneous telemetry ingestion and downstream inference. To play this role reliably, it must define:
\begin{enumerate}
    \item a strict canonical core over entities, relationships, state, and provenance;
    \item explicit rules for identity persistence and temporal validity;
    \item extensible mapping zones for source-specific nuance; and
    \item stable learning-facing views that downstream AI systems can consume without rebuilding telemetry semantics from scratch.
\end{enumerate}

\subsection{Canonical Entities}

Let $\mathcal{E}$ denote the finite set of core canonical entity types:
\begin{equation}
\begin{aligned}
\mathcal{E} = \{&
\texttt{Host},\,
\texttt{User},\,
\texttt{Process},\,
\texttt{File}, \\
&
\texttt{NetworkFlow},\,
\texttt{CloudResource}, \\
&
\texttt{Credential},\,
\texttt{ExternalEntity}
\}.
\end{aligned}
\end{equation}

These types form the strict core of the CSTS entity layer. They are intentionally broad enough to absorb common enterprise telemetry domains while remaining semantically stable across sources. Additional domain-specific entity types may be introduced through governed extensions, but downstream portability depends on maintaining a small, stable canonical core.

Each canonical entity $e \in V_t$ is defined as
\begin{equation}
e = (id, type, A_e, S_e, T_e),
\end{equation}
where:
\begin{itemize}
    \item $id$ is a globally unique canonical identifier;
    \item $type \in \mathcal{E}$;
    \item $A_e$ is a normalized attribute map;
    \item $S_e$ records source metadata, mapping lineage, and confidence;
    \item $T_e = [t_{\mathrm{start}}, t_{\mathrm{end}}]$ is the temporal validity interval.
\end{itemize}

\subsubsection*{Entity invariants}

All canonical entities must satisfy the following invariants:
\begin{enumerate}
    \item \textbf{Stable identity:} $id$ remains invariant across ingestion boundaries once canonicalized.
    \item \textbf{Schema normalization:} attributes conform to canonical types and governed naming conventions.
    \item \textbf{Temporal explicitness:} validity intervals are explicit rather than inferred only downstream.
    \item \textbf{Provenance preservation:} source mappings, adapter lineage, and confidence metadata are retained.
\end{enumerate}

These invariants distinguish CSTS from ordinary event normalization. The substrate does not simply rename fields; it creates canonical objects whose identity and temporal meaning are stable enough to support downstream graph reasoning, statistical modeling, and representation learning.

\subsubsection*{Identity resolution}

Let $\mathcal{O}$ denote raw telemetry observations. CSTS defines a resolution function
\begin{equation}
R: \mathcal{O} \rightarrow id
\end{equation}
such that semantically equivalent observations map to the same canonical identity:
\begin{equation}
R(o_i) = R(o_j) \Rightarrow o_i \sim o_j.
\end{equation}
Resolution may be deterministic (e.g., normalized identifiers), probabilistic (record linkage), or graph-based (co-reference clustering). In all cases, merges must preserve lineage, confidence, and auditability.

This is a key architectural point: identity reconciliation is a substrate function, not a model-specific convenience step. By moving identity stabilization into CSTS, downstream systems inherit stable object identity rather than repeatedly reimplementing stitching logic under different vendor and deployment assumptions.

\subsubsection*{Lifecycle model}

Entities transition through time-indexed states:
\begin{equation}
\texttt{Created} \rightarrow \texttt{Active} \rightarrow \texttt{Dormant} \rightarrow \texttt{Retired}.
\end{equation}
Lifecycle transitions are immutable once recorded and support replay, audit, and historical reasoning. This matters especially for enterprise AI because many learning systems depend not only on the current existence of an entity, but on its persistence, inactivity periods, and reactivation history.

\subsection{Canonical Relationships}

Let $\mathcal{R}$ denote the finite set of core canonical relationship types:
\begin{equation}
\begin{aligned}
\mathcal{R} = \{&
\texttt{AUTHENTICATES\_TO},\,
\texttt{EXECUTES}, \\
&
\texttt{CONNECTS\_TO},\,
\texttt{READS}, \\
&
\texttt{WRITES},\,
\texttt{MODIFIES}, \\
&
\texttt{SPAWNS},\,
\texttt{OWNS}, \\
&
\texttt{ASSOCIATED\_WITH}
\}.
\end{aligned}
\end{equation}

These are the core relationship semantics required to express a large fraction of operational cyber behavior across endpoint, network, identity, and cloud telemetry. As with entities, the goal is not to enumerate every possible source-level relation in advance, but to define a stable canonical core with governed extension paths.

Each relationship $r \in E_t$ is defined as
\begin{equation}
r = (src, dst, type, A_r, t_r, P_r),
\end{equation}
where:
\begin{itemize}
    \item $src, dst \in V_t$ are canonical entities;
    \item $type \in \mathcal{R}$;
    \item $A_r$ is a relationship attribute map;
    \item $t_r$ is a timestamp or interval;
    \item $P_r$ encodes provenance metadata.
\end{itemize}

\subsubsection*{Structural constraints}

CSTS enforces structural constraints to ensure semantic consistency:
\begin{enumerate}
    \item \textbf{Type consistency:} relationship domains and codomains are fixed or governed (e.g., $\texttt{AUTHENTICATES\_TO(User, Host)}$).
    \item \textbf{Directionality:} all edges are directed.
    \item \textbf{Temporal alignment:} $t_r \subseteq T_{src} \cap T_{dst}$ whenever the relationship is asserted between temporally valid entities.
    \item \textbf{Cardinality rules:} optional multiplicity constraints may be enforced for governed relation families.
\end{enumerate}

\subsubsection*{Provenance requirements}

Each relationship includes provenance sufficient for auditability and explainability, including:
\begin{itemize}
    \item source system identifier,
    \item ingestion timestamp,
    \item confidence score,
    \item transformation lineage,
    \item mapping profile version.
\end{itemize}

This requirement is essential because CSTS is intended to preserve source-specific nuance without allowing that nuance to leak unpredictably into downstream interfaces.

\subsection{Temporal Graph Evolution}

Security telemetry is dynamic. CSTS models graph evolution incrementally:
\begin{equation}
G_{t+1} = G_t + \Delta V_t + \Delta E_t,
\end{equation}
where $\Delta V_t$ and $\Delta E_t$ denote entity and relationship updates, respectively.

This notation is intentionally simple, but conceptually important. CSTS is not a static graph export; it is a temporally evolving canonical substrate. This allows downstream systems to operate over state transitions, behavioral windows, and historical context without reconstructing temporal meaning from disconnected event records.

\subsubsection*{Time semantics}

CSTS distinguishes multiple notions of time:
\begin{itemize}
    \item \textbf{event time} ($t_e$): when the source event occurred;
    \item \textbf{ingestion time} ($t_i$): when the observation entered the substrate;
    \item \textbf{valid time} ($t_v$): when the represented state or relationship is considered true in the modeled world.
\end{itemize}
Attributes may therefore be time-indexed:
\begin{equation}
A_e(t), \qquad A_r(t).
\end{equation}

This separation is important for replay, delayed ingest, backfill, audit, and learning tasks over longitudinal behavior. In many operational systems, event time and ingestion time diverge materially; CSTS preserves that distinction explicitly.

\subsubsection*{Canonical aggregation}

Aggregation operators are defined over canonical entities and relationships rather than raw vendor events. Supported abstractions include sliding windows, tumbling windows, sessionization, motif extraction, and cross-entity temporal joins. This design ensures that aggregation semantics remain stable across heterogeneous telemetry sources. A model consuming a 24-hour host-state window or a bounded authentication subgraph should see the same canonical object class regardless of which vendor feeds generated the underlying raw records.

\subsection{Learning Objects and Representation Views over CSTS}
\label{sec:csts_learning_objects}

Although CSTS is defined as a time-indexed attributed multigraph over canonical entities and typed relationships, downstream learning systems typically do not consume the full substrate directly. Instead, they operate on bounded relational-temporal units derived from the substrate. To make this interface explicit, we define a \emph{learning object} in CSTS as a semantically governed, time-bounded view over canonical entities, relationships, state, and provenance used as the input unit for downstream statistical, graph, or representation-learning models.

This extension is not a separate ontology. CSTS remains ontologically entity-relational. Learning objects define the \emph{model-consumption contract} over existing CSTS primitives.

Formally, let
\[
G_t = (V_t, E_t)
\]
denote the CSTS graph at time $t$. A CSTS learning object is a tuple
\[
o = (V_o, E_o, T_o, X_o, P_o, \phi_o),
\]
where:
\begin{itemize}
    \item $V_o \subseteq V_t$ is a bounded set of canonical entities;
    \item $E_o \subseteq E_t$ is a bounded set of typed relationships among those entities;
    \item $T_o = [t_a, t_b]$ is a bounded temporal support interval;
    \item $X_o$ is the set of canonical attributes and derived behavioral summaries over $T_o$;
    \item $P_o$ records provenance, adapter lineage, and confidence metadata inherited from the substrate;
    \item $\phi_o$ is an optional focalization map specifying the central entity, interaction, path, or motif around which the object is constructed.
\end{itemize}

Several classes of learning objects arise naturally in CSTS.

\paragraph{1) Entity-state objects.}
An \emph{entity-state object} represents the temporally indexed behavioral state of a single canonical entity over a bounded interval. A host-state object may summarize authentication neighborhood, process activity, file-access diversity, and network role over a fixed window. A user-state object may summarize login cadence, peer-host diversity, privilege spread, and role deviation relative to historical baselines. These objects are natural for identity-centric analytics such as lateral movement detection, insider-threat modeling, and behavioral risk scoring.

\paragraph{2) Interaction-state objects.}
An \emph{interaction-state object} represents a focal typed interaction together with its source-target context. For example, a focal \texttt{CONNECTS\_TO} edge between a process or host and an external destination may be enriched with prior state, destination neighborhood, recent authentication activity, associated credentials, and bounded graph statistics over interval $T_o$. This object class is useful for flow-centric anomaly detection, zero-day detection, and edge-level scoring tasks.

\paragraph{3) Subgraph-state objects.}
A \emph{subgraph-state object} represents a bounded typed neighborhood around a focal entity or interaction, typically at one-hop or two-hop depth, together with explicit temporal support. Such objects preserve local relational structure without requiring the full enterprise graph to be processed at once. They are appropriate for graph neural networks, neighborhood aggregation methods, motif-sensitive detectors, and hybrid graph-tabular encoders.

\paragraph{4) Motif-state objects.}
A \emph{motif-state object} represents a recurrent local behavioral pattern expressed over canonical entities and typed relationships. Examples include fan-out authentication bursts, process-to-file write cascades, privilege-boundary traversals, repeated remote-execution chains, and short-window bipartite expansion patterns consistent with coordinated attack behavior. The purpose of this class is to elevate repeated relational structure to a first-class learning unit rather than forcing models to rediscover it from raw events each time.

\paragraph{Representation views.}
To support modern representation learning, we define a \emph{view} of a CSTS learning object as a semantics-preserving transformation
\[
v : o \mapsto \widetilde{o}
\]
such that $\widetilde{o}$ preserves the canonical identity, typed relational interpretation, temporal validity, and provenance requirements of the underlying object while varying its presentation to a downstream model.

Examples of valid CSTS views include:
\begin{itemize}
    \item \textbf{attribute-masked views:} selected non-critical attributes in $X_o$ are hidden or perturbed while preserving core semantics;
    \item \textbf{neighborhood-subsampled views:} a bounded subset of a local neighborhood is sampled while preserving focal identity and core typed relations;
    \item \textbf{adjacent-window views:} temporally nearby slices of the same object are formed to capture behavioral continuity;
    \item \textbf{partial-source views:} one telemetry producer or modality is omitted while preserving canonical object meaning;
    \item \textbf{modality-restricted views:} tabular-only, temporal-only, graph-only, or hybrid projections derived from the same underlying object.
\end{itemize}

A valid CSTS view must satisfy:
\begin{enumerate}
    \item \textbf{identity preservation:} focal canonical identity remains unchanged;
    \item \textbf{relational validity:} typed relationships remain semantically valid under domain/codomain constraints;
    \item \textbf{temporal consistency:} the transformed object preserves a coherent temporal support interval or admissible sub-interval;
    \item \textbf{provenance traceability:} lineage, source metadata, and confidence annotations remain available;
    \item \textbf{semantic admissibility:} the transformation may reduce observability or alter resolution, but it must not change the underlying behavioral interpretation.
\end{enumerate}

These constraints connect representation learning to the portability goals of CSTS. The substrate is intended to provide schema stability and canonicalization guarantees at the telemetry layer; learning-object and view semantics extend that stability upward into model construction. In this sense, CSTS does not merely normalize telemetry. It defines a governed representational basis for substrate-native anomaly detection, novelty detection, metric learning, contrastive learning, graph-temporal reasoning, and future AI systems over persistent security state \cite{zdt_flow,zdt_metric,koukoulis2025selfsup,wu2024egconmix,wilkie2026contrastive,khosla2020supcon}.

\subsection{Schema Governance and Evolution}

Long-term portability requires explicit schema governance. Each substrate release is versioned:
\begin{equation}
\texttt{Schema}_v.
\end{equation}
Adapters and mapping profiles declare compatibility with specific versions.

\subsubsection*{Core-versus-extension discipline}

CSTS distinguishes between:
\begin{itemize}
    \item a \textbf{strict core}, consisting of canonical entity types, relation types, time semantics, and provenance requirements on which downstream portability depends; and
    \item an \textbf{extension zone}, in which source-specific residuals, custom attributes, and future telemetry specializations may be preserved without altering the canonical downstream interface.
\end{itemize}

This distinction is critical. A substrate that is too rigid cannot absorb real-world telemetry; a substrate that is too permissive cannot stabilize inference. CSTS therefore governs extensibility explicitly rather than leaving it implicit.

\subsubsection*{Backward-compatibility rules}

\begin{enumerate}
    \item Additive attribute extensions are permitted.
    \item Renaming requires alias mapping and version annotation.
    \item Removal of required fields mandates deprecation periods.
    \item Core identity, relationship, temporal, and provenance invariants must not be broken by schema evolution.
\end{enumerate}

\subsection{Representational Guarantees}

CSTS enforces the following representational guarantees at the telemetry layer:
\begin{enumerate}
    \item persistent canonical identity across ingestion boundaries;
    \item environment-independent relationship typing under governed semantics;
    \item explicit temporal continuity and time semantics;
    \item provenance-preserving canonicalization;
    \item versioned schema governance with stable downstream contracts.
\end{enumerate}

These guarantees stabilize the representational boundary between telemetry ingestion and downstream modeling. CSTS does not guarantee downstream model generalization by itself; rather, it isolates portability constraints to the representation layer so that modeling-level properties can be analyzed independently.

\subsection*{Implications}

This section formalized CSTS as a time-indexed, entity-centric multigraph with explicit identity, relationship, temporal, provenance, and governance invariants. By elevating security telemetry from raw event streams to stable canonical entities and relationships, CSTS provides a structurally consistent substrate for downstream analytics. The formal definitions ensure that cross-source integration, temporal reasoning, provenance preservation, and stable learning-object construction are enforced at the architectural level rather than deferred to model-specific preprocessing.

Crucially, CSTS shifts the stability requirement from machine-learning pipelines to the telemetry substrate itself. Rather than compensating for drift, schema fragmentation, and identity inconsistency through repeated retraining or source-specific heuristics, CSTS defines structural guarantees that persist across heterogeneous deployments. In doing so, it reframes a large part of cybersecurity AI fragility as a representational problem with a systematic architectural solution.

%%%%%%%%%%%%%%%%%%%%%%%%%%%%%%%%%%%%%%%%%%%%%%%%%
\section{Case Studies}

Security analytics systems often fail in deployment not because the underlying detection ideas are weak, but because the surrounding telemetry substrate is unstable. Lateral movement detectors, anomaly models, ransomware analytics, and graph-based reasoning systems are frequently rebuilt for each environment through source-specific parsing, identity stitching, relationship extraction, and feature engineering. The result is that the same nominal detection task is repeatedly re-instantiated over different telemetry semantics, making portability fragile and deployment expensive.

CSTS addresses this problem by separating \emph{representation} from \emph{detection logic}. Once telemetry has been mapped into persistent canonical entities, typed relationships, temporal state, and provenance-preserving views, downstream methods can operate over a stable substrate rather than over brittle vendor-specific event layouts. The purpose of the present section is therefore not to introduce new detection algorithms. Rather, it is to show that several well-known cybersecurity detection problems can be reinterpreted cleanly through the CSTS lens. This illustrates the broader claim of the paper: once telemetry is expressed as canonical interactions over persistent entities, seemingly distinct security tasks reduce to identifying structured deviations in an evolving entity-relational state space.

\subsection{Lateral Movement Under CSTS}

Lateral movement detection has traditionally relied on authentication sequences, credential reuse patterns, remote session behavior, and multi-host traversal analysis. Prior work demonstrates strong performance using user-behavior features engineered from authentication telemetry~\cite{lateral_movement_uba}. However, these pipelines typically reconstruct identity and interaction structure implicitly, and their feature logic is often tightly coupled to environment-specific log semantics.

Within CSTS, lateral movement is more naturally interpreted as \emph{role deviation and traversal emergence} in a constrained entity-relational graph. Canonical entities such as \texttt{User}, \texttt{Host}, \texttt{Credential}, and \texttt{Process} are connected through typed relationships including \texttt{AUTHENTICATES\_TO}, \texttt{EXECUTES}, and \texttt{SPAWNS}. Under this view, lateral movement is not primarily a sequence of isolated login events; it is the appearance of traversal paths that are inconsistent with the historical role geometry of a user, credential, or host neighborhood.

Signals described in~\cite{lateral_movement_uba}---such as unusual host adjacency, abnormal session behavior, or deviations from typical login rhythms---map naturally to CSTS-level graph deviations: novel \texttt{User}$\rightarrow$\texttt{Host} adjacencies, rapid ego-network expansion, unexpectedly persistent credential-host pathways, or traversal chains that exceed typical path-length and role-transition patterns. Because identity and relationship semantics are stabilized at the substrate layer, the detection problem becomes one of structural and behavioral deviation over canonical objects rather than one of vendor-specific authentication field engineering. This is precisely the kind of shift CSTS is intended to enable.

\subsection{Zero-Day Threat Detection Under CSTS}

Zero-day detection is often formulated as anomaly detection over network telemetry augmented with structural context~\cite{zdt_flow}, or as a distinction between anomaly detection and novelty detection using metric-learning or autoencoder-based representations~\cite{zdt_metric}. These formulations all depend, implicitly or explicitly, on learning a notion of normal behavior and identifying previously unseen deviations from that learned structure.

Under CSTS, zero-day detection can be interpreted as identifying \emph{relational deformation} in the geometry of canonical entities and interactions. In graph-and-flow approaches~\cite{zdt_flow}, network flows are augmented with topological features such as neighborhood structure, centrality, or path context. In CSTS, these structural properties arise directly from the substrate, because persistent identity and typed relationships induce a canonical interaction graph rather than requiring graph construction to be reinvented for each dataset or deployment.

From this perspective, anomaly detection corresponds to identifying interactions or bounded relational objects that cannot be reconstructed under learned substrate-consistent behavioral geometry, aligning naturally with reconstruction-based or autoencoder-like approaches~\cite{sakurada2014anomaly}. Novelty detection corresponds to metric or manifold separation between previously observed canonical interaction patterns and new relational configurations~\cite{zdt_metric}. The practical benefit is that learning shifts away from raw identifiers, vendor field names, or source-specific parser artifacts and toward the behavior of canonical objects. Inference is therefore grounded more directly in interaction structure and less in telemetry surface syntax.

\subsection{Ransomware as Motif Emergence}

Ransomware detection is often based on file entropy signals, process bursts, or signature-like heuristics. Within CSTS, ransomware can be reinterpreted more generally as the emergence of a characteristic relational motif in the canonical entity graph.

A canonical ransomware pattern typically includes execution of an anomalous or low-prevalence binary, followed by rapid and high-density \texttt{WRITES} or \texttt{MODIFIES} relationships between a process entity and many file entities within short temporal windows, often accompanied by privilege-boundary crossing, process spawning behavior, or unusual host-level state changes. In graph-theoretic terms, this appears as abrupt degree expansion and repeated bipartite interaction between a focal process node and a large file neighborhood.

Cross-temporal ransomware research has emphasized the need to detect campaigns under changing conditions rather than relying only on static signatures~\cite{ransomware_alert_graph}. CSTS supports this objective by stabilizing the identity of the acting process, the file neighborhood, the relevant typed relationships, and the temporal support of the motif across deployments. The result is that detection can focus more directly on repeated structural replication patterns and motif emergence~\cite{milo2002network} rather than on source-specific file system field conventions. This is not merely a graph convenience; it is an example of how a canonical substrate can turn otherwise source-bound heuristics into portable behavioral objects.

\subsection{Implications of the Case Studies}

Taken together, these case studies illustrate a common theme: lateral movement, zero-day activity, and ransomware can all be expressed as structured deviations in a persistent entity-relational substrate. Their surface telemetry differs, and the downstream models used to score them may differ, but the underlying representational need is the same. Each task requires stable identity, typed relationship semantics, temporal continuity, and a bounded object of analysis over which detection or learning can operate.

This is the deeper role of CSTS. It does not make all downstream modeling problems trivial, and it does not guarantee cross-environment generalization by itself. What it does is move the burden of identity stabilization, relation semantics, and time-aware object construction out of individual pipelines and into a shared substrate. That shift clarifies the boundary between representational stability and model-level generalization. Once telemetry is aligned to canonical entities and typed relationships, downstream systems can focus more directly on the real problem: distinguishing benign and malicious behavior over stable semantic objects rather than compensating for telemetry fragmentation.

In this sense, the case studies are not merely examples. They are evidence that a wide range of cybersecurity tasks can be re-expressed over the same canonical substrate. This is the architectural basis on which CSTS aims to support portable anomaly detection, graph reasoning, behavioral modeling, and future AI-native security workflows across heterogeneous enterprise environments.

%------------------------------------------------
\section{Deployment Model and Tiered Adoption}

CSTS is designed for incremental adoption rather than wholesale infrastructure replacement. Enterprise security environments vary significantly in maturity, telemetry quality, deployment topology, governance requirements, and integration capability. A viable telemetry substrate must therefore support staged deployment, allowing organizations to progressively increase semantic richness without disrupting existing pipelines, SIEM workflows, data lake infrastructure, or model-serving systems.

This incremental view is especially important because enterprise deployments occur across cloud, on-prem, and hybrid environments. Some organizations begin with vendor-normalized logs in a SIEM; others already maintain lakehouse-style telemetry pipelines; still others operate mixed ecosystems across multiple clouds and legacy on-prem infrastructure. CSTS is intended to function across these settings by preserving a stable canonical core while allowing source-specific and deployment-specific variation to remain at the adapter and orchestration layers. In this sense, the adoption model is not merely a maturity ladder; it is also a portability model.

We therefore define a five-tier adoption framework that enables organizations to move from basic event normalization toward a fully AI-ready, embedding-compatible, entity-relational telemetry substrate.

\subsection{Tier 0: Event Normalization}

Tier 0 corresponds to alignment with existing telemetry normalization frameworks such as OCSF~\cite{ocsf}, Elastic Common Schema (ECS)~\cite{ecs}, and OpenTelemetry~\cite{opentelemetry}. At this stage, organizations standardize field names, event categories, and log formats across vendors or telemetry producers.

Normalization improves interoperability, reduces parsing overhead, and enables more consistent ingestion into SIEM, observability, and data lake environments. However, it remains event-centric. Entities are not persistently resolved, relationships remain implicit, temporal behavioral state is not yet stabilized, and provenance-aware canonical object construction is deferred downstream. Tier~0 therefore reduces syntactic inconsistency but does not yet address the representational stability required for portable AI systems. CSTS treats Tier~0 as compatible and often necessary, but insufficient as a final substrate.

\subsection{Tier 1: Entity Resolution}

Tier 1 introduces persistent canonical identity across telemetry sources. Raw observations are mapped to canonical entity types such as \texttt{User}, \texttt{Host}, \texttt{Process}, \texttt{Credential}, and \texttt{CloudResource}, each with a globally stable identifier. This tier implements the canonical resolution function defined in the formal substrate specification, ensuring that semantically equivalent observations across heterogeneous systems map to a consistent identity.

Entity resolution is the minimal viable CSTS substrate. Once identity persistence is enforced, downstream analytics can operate over stable actors and objects rather than over ephemeral records. This addresses one of the most common causes of deployment fragility in AI-based security systems: the fact that the same logical user, host, or workload may appear under inconsistent identifiers across vendors, time periods, and environments. By moving identity stabilization into the substrate, CSTS begins to decouple downstream inference from source-specific identifier instability.

Tier~1 also has a practical adoption advantage: it can often be introduced without changing downstream model logic immediately. Existing systems may continue consuming familiar feature pipelines while benefitting from more stable canonical identity underneath.

\subsection{Tier 2: Relationship Materialization}

Tier 2 materializes typed relationships among canonical entities, forming an explicit entity-relational graph. Authentication, process execution, file modification, network flow, service access, and cloud interaction events are transformed into directed typed edges with governed domain and codomain constraints.

This tier aligns directly with graph-based detection paradigms demonstrated in prior work on lateral movement, anomaly detection, and zero-day threat modeling~\cite{lateral_movement_uba,zdt_flow}. Instead of reconstructing relationships independently inside each detection pipeline, CSTS defines them once at the substrate level and exposes them as reusable canonical objects.

Materializing relationships changes the role of telemetry fundamentally. At Tier~0 and Tier~1, telemetry is still primarily normalized data. At Tier~2, it becomes a reusable analytical substrate. Graph structure, motif emergence, interaction concentration, and role-transition logic can now be derived from stable canonical relations rather than from source-specific parser logic. This reduces repeated feature engineering, improves reproducibility, and allows multiple downstream systems to operate over the same relational semantics.

\subsection{Tier 3: Temporal Modeling}

Tier 3 incorporates explicit temporal state modeling into the substrate. Entities and relationships are represented as time-indexed objects with validity intervals, rolling aggregation windows, state updates, and support for temporal joins. This enables behavioral baselining, sessionization, sequence extraction, and longitudinal role analysis at the substrate layer rather than within model-specific code.

Temporal modeling is critical for anomaly detection, novelty detection, forecasting, and behavior-based reasoning~\cite{zdt_metric}. Without explicit temporal semantics, models must infer continuity from fragmented logs or rebuild time-aware objects on a per-pipeline basis. CSTS instead enforces temporal explicitness so that rolling behavioral state is represented consistently across environments.

Tier~3 is also where CSTS begins to support richer dynamical analyses, including bounded behavioral windows, state trajectories, and regime-aware inference. Even when the downstream model is not explicitly graph-based, consistent temporal state reduces representational variability and improves cross-environment comparability of behavioral abstractions. This does not guarantee generalization, but it reduces the share of deployment failure attributable to unstable temporal semantics.

\subsection{Tier 4: AI-Ready and Embedding-Compatible Extensions}

Tier 4 extends CSTS from a stable entity-relational substrate into an explicitly AI-ready inference layer. At this stage, canonical entities, relationships, and temporal objects expose governed feature and learning-object contracts suitable for graph embedding, contrastive learning, metric learning, sequence modeling, autoencoder-based anomaly detection, and related AI-native tasks.

Embedding-compatible extensions do not replace earlier tiers; they build upon stable identity, typed relationships, and temporal state. Prior work demonstrates the effectiveness of metric learning and representation learning for zero-day detection and related novelty-sensitive tasks~\cite{zdt_metric,martinez2024metricfused,koukoulis2025selfsup}. However, such models are often coupled to source-specific features and bespoke preprocessing logic. CSTS changes this by defining learning-facing interfaces over canonical objects rather than over raw telemetry schemas.

At Tier~4, the substrate becomes a true inference boundary. Downstream systems can consume stable learning objects, graph neighborhoods, temporal windows, and representation views derived from canonical entities and relationships rather than from vendor-specific event artifacts. This is the point at which CSTS supports not only structural normalization, but also plug-and-play AI workflows: anomaly detection, novelty detection, motif reasoning, graph learning, retrieval, and future agentic systems can all operate over a shared canonical substrate.

\subsection{Operational Interpretation of the Tier Model}

The five-tier model is intended to be operational, not merely conceptual. Organizations need not implement all tiers at once. A realistic deployment path may begin with Tier~0 normalization already in place, add Tier~1 identity persistence and Tier~2 relation materialization for one or two high-value use cases, then extend to Tier~3 temporal modeling and Tier~4 AI-ready views as the telemetry platform matures.

This staged approach supports practical adoption across enterprise realities:
\begin{itemize}
    \item \textbf{Cloud-first deployments} may begin by canonicalizing cloud-native logs into CSTS and then layering richer identity and relation models over time.
    \item \textbf{On-prem deployments} may begin with normalized SIEM exports and evolve toward CSTS-backed graph and behavioral analytics without rearchitecting all infrastructure at once.
    \item \textbf{Hybrid environments} may use CSTS as the common substrate across cloud and on-prem telemetry while keeping ingestion adapters deployment-specific.
\end{itemize}

Thus, the tier model is also a deployment-decoupling model: the environment-specific edge can vary while the canonical substrate and downstream AI interfaces remain stable.

\subsection{Summary}

In combination, these tiers define a progression from syntactic normalization to stable AI-native telemetry representation. Tier~0 addresses event-level consistency. Tier~1 stabilizes identity. Tier~2 materializes relationships. Tier~3 introduces explicit temporal state. Tier~4 exposes learning-compatible and inference-ready interfaces over the resulting canonical substrate.

The significance of this progression is architectural. CSTS is not intended to be adopted all at once or only in its most expressive form. Rather, it provides a governed path by which enterprises can move from fragmented telemetry toward a canonical, portable, and AI-ready representation without requiring wholesale replacement of existing systems. This staged adoption model is one of the main reasons CSTS is practical: it turns a difficult representational transition into an incremental engineering path.

%%%%%%%%%%%%%%%%%%%%%%%%%%%%%%%%%%%%%%%%%%%%%%%%%%%%%%%%%%%%%%%%%%%%%%%%%%%%%%%%
\section{Portability Decomposition Framework}
\label{sec:portability}

CSTS is designed to stabilize telemetry representation across heterogeneous enterprise environments. However, representational stability alone does not guarantee model portability. A model may still fail after telemetry has been successfully canonicalized, and it is therefore essential to distinguish \emph{where} portability breaks. To reason precisely about deployment behavior, we decompose portability into three distinct layers: schema stability, representational invariance, and semantic orientation stability.

This decomposition is important because portability failures are often conflated in practice. A detector may fail because a field changed name, because identity or relationship semantics drifted across environments, or because the learned feature geometry itself no longer aligns with the target distribution. These are different failure modes and require different remedies. CSTS is intended to eliminate or sharply reduce the first two by enforcing a stable substrate. The third remains a property of the learning problem itself.

\subsection{Schema Stability}

\paragraph{Definition (Schema Stability).}
A detection pipeline satisfies \emph{schema stability} if feature definitions, object construction rules, and aggregation logic remain unchanged under field renaming, deletion, format variation, parser change, or vendor-specific schema drift.

Schema instability appears when upstream telemetry modifications break ingestion, feature extraction, graph construction, or preprocessing pipelines. Event-centric normalization frameworks such as OCSF~\cite{ocsf} reduce syntactic inconsistency, but they do not by themselves guarantee that downstream pipelines remain stable under source evolution. In many environments, the same conceptual telemetry signal can still arrive under modified field names, altered producer conventions, or partial source loss, forcing repeated downstream repair.

CSTS addresses this by requiring that raw telemetry variation be absorbed at the adapter and mapping layer. The stable downstream interface is not the raw event schema, but the canonical entity-relational substrate. In that sense, schema stability is the outermost portability layer: if it fails, downstream portability cannot even be meaningfully evaluated because the analytic object itself is not preserved.

\subsection{Representational Invariance}

\paragraph{Definition (Representational Invariance).}
A telemetry substrate satisfies \emph{representational invariance} if:
\begin{enumerate}
    \item persistent entity identity is maintained across ingestion boundaries,
    \item relationship type semantics are environment-independent under governed mappings,
    \item temporal ordering and state continuity are preserved.
\end{enumerate}

Representational invariance is stronger than schema stability. Schema stability ensures that pipelines continue to run; representational invariance ensures that what they run \emph{means the same thing}. A telemetry system may preserve field names and still fail representationally if users, hosts, credentials, processes, or network interactions are reconstructed inconsistently across environments. Likewise, graph structure may remain syntactically valid while its semantic interpretation drifts because relationship extraction rules are environment-specific or poorly governed.

CSTS enforces representational invariance through canonical identity resolution, typed relationship constraints, explicit temporal modeling, provenance-preserving mappings, and governance-backed schema evolution. The substrate therefore guarantees that downstream systems see stable canonical objects rather than source-specific encodings. This does not mean that all environments become identical; rather, it means that structural comparability is preserved at the telemetry layer.

Importantly, representational invariance guarantees comparability of entities, relations, and time-indexed state across environments, but it does not guarantee that learned features or models behave identically under domain shift. That stronger property belongs to the next layer.

\subsection{Semantic Orientation Stability}

Even when schema stability and representational invariance hold, learned features may invert, weaken, or degrade under distribution shift. This is not a telemetry-ingestion failure; it is a modeling-level portability failure.

\paragraph{Definition (Semantic Orientation Stability).}
A feature $f$ exhibits \emph{semantic orientation stability} between environments $\mathcal{E}_A$ and $\mathcal{E}_B$ if the sign of its class-conditional separation remains invariant:
\begin{equation}
\mathrm{sign}\!\big(\Delta_A(f)\big)
=
\mathrm{sign}\!\big(\Delta_B(f)\big),
\end{equation}
where
\[
\Delta_\mathcal{E}(f)
=
\mathbb{E}_{\mathcal{E}}[f \mid y=1]
-
\mathbb{E}_{\mathcal{E}}[f \mid y=0].
\]

If the sign reverses, a feature that was positively associated with malicious behavior in one environment becomes negatively associated in another. More generally, even without full sign reversal, substantial attenuation or distortion of $\Delta_\mathcal{E}(f)$ may degrade transfer. This phenomenon reflects instability in the behavioral meaning or discriminative geometry of the feature under domain shift rather than instability in telemetry ingestion or canonicalization.

This distinction is central to the CSTS thesis. The purpose of CSTS is not to guarantee downstream model success in every environment. Rather, it is to isolate portability failures so that one can determine whether the failure is caused by broken telemetry representation or by genuine distributional and modeling differences.

\subsection{Portability Failure Modes}

The decomposition above yields three distinct classes of portability failure:
\begin{enumerate}
    \item \textbf{Schema failure:} feature extraction or preprocessing breaks because upstream fields drift, disappear, or change form.
    \item \textbf{Representational failure:} entity identity, typed relationships, provenance, or temporal semantics are inconsistent across environments.
    \item \textbf{Orientation failure:} learned feature direction or discriminative geometry changes under domain shift despite stable representation.
\end{enumerate}

These regimes should not be conflated. Schema failure is primarily an ingestion and interface problem. Representational failure is a substrate problem. Orientation failure is a learning and data-distribution problem. The practical importance of the decomposition is that different mitigation strategies apply to each one. Adapter redesign and schema governance address the first. Canonical identity, typed relations, and temporal state address the second. Model redesign, recalibration, domain adaptation, or task reformulation address the third.

CSTS is designed to eliminate or sharply reduce the first two categories by enforcing structural invariants at the telemetry layer. Residual degradation after successful CSTS alignment should therefore be interpreted more honestly: not as an ingestion failure, but as evidence of modeling-level semantic instability or distribution shift.

\subsection{Implications for Detection Tasks}

This framework helps explain why different detection paradigms exhibit different portability behavior.

Identity-centric tasks such as lateral movement detection depend heavily on stable relational topology and persistent identity semantics. When canonical users, hosts, credentials, and typed authentication relationships are preserved, the behavioral meaning of graph-derived features often remains relatively stable across environments. In such tasks, a large share of deployment fragility can be attributed to schema and representational instability, which CSTS is explicitly designed to address.

By contrast, anomaly- and geometry-based tasks such as zero-day detection often depend more strongly on global distributional properties of interaction structure, neighborhood geometry, or latent manifold shape. Even under schema stability and representational invariance, these properties may still shift across enterprise topologies and telemetry regimes. In those cases, residual degradation reflects semantic orientation instability rather than a failure of the canonical substrate.

This distinction directly motivates the experimental evaluation in the paper. The central empirical question is not merely whether CSTS improves performance. It is whether CSTS removes schema and representational failures so that remaining degradation can be identified as a modeling-level portability boundary rather than as a consequence of telemetry fragmentation. That is a stronger and more actionable result: it turns vague deployment fragility into a structured diagnostic framework.

%-------------------------------------------
\section{Experimental Validation}

This section evaluates CSTS as a representational substrate for portable detection across heterogeneous enterprise environments. The goal is not simply to report task accuracy, but to test whether the structural guarantees formalized earlier translate into measurable deployment benefits. In particular, we evaluate three stability properties derived from Sections~\ref{sec:formal_definition} and~\ref{sec:portability}: representational invariance, schema stability, and semantic orientation stability.

Accordingly, we test the following hypotheses:

\textbf{H1 (Representational Invariance).}
Enforcing canonical entity persistence and typed relational semantics improves cross-topology generalization for identity-centric detection tasks.

\textbf{H2 (Schema Stability).}
Confining raw-schema variability to thin canonical adapter layers prevents detection collapse under field-level perturbation.

\textbf{H3 (Semantic Orientation Boundary).}
Representational invariance does not guarantee semantic orientation stability under domain shift.

We study these hypotheses using two main task families. The first is lateral movement (LM) detection derived from authentication and process telemetry~\cite{lateral_movement_uba}, which is naturally identity-centric and therefore a strong test of entity persistence and typed relationship semantics. The second is flow-centric zero-day threat detection (ZDT)~\cite{zdt_flow,zdt_metric}, which places more weight on global structural and distributional geometry and therefore serves as a natural test of semantic orientation stability under domain shift.

Throughout, we hold model architecture, windowing, training splits, hyperparameters, and random seeds fixed across pipelines whenever possible. This is critical: the purpose of the evaluation is to isolate representational effects rather than to compare algorithm families.

\subsection{Evaluation Setup}

We construct two enterprise environments:
\begin{itemize}
    \item \textbf{Env A}: source training environment;
    \item \textbf{Env B}: target deployment environment with schema variation, identifier differences, and increased benign behavioral diversity.
\end{itemize}

For each environment, we compare two pipelines.

\paragraph{Event-centric baseline.}
Feature extraction operates directly on raw telemetry fields (e.g., user identifiers, destination hosts, timestamps, and event-type labels). Features consist primarily of windowed count, cardinality, and source-specific aggregate statistics.

\paragraph{CSTS pipeline.}
Raw telemetry is first mapped into canonical entities and typed relationships. Windowed features are then derived from canonical identifiers, relational structure (e.g., unique destination hosts, neighborhood expansion), and training-history-based rarity measures computed strictly on the training split to prevent leakage.

Both pipelines use identical 30-minute windows and identical classifier configurations. Hyperparameters, thresholds, and training seeds are held constant across experiments so that performance differences reflect representational properties rather than algorithmic variation. In addition to F1 at threshold 0.5, we report AUROC and best-F1 obtained via a fixed threshold sweep to separate representation quality from calibration artifacts.

\subsection{Controlled Transfer Benchmark (Synthetic EnvA/EnvB)}
\label{sec:controlled_lm}

We first evaluate CSTS in a controlled cross-topology benchmark designed to isolate representational effects from labeling sparsity, uncontrolled logging artifacts, and schema incompleteness. Unlike the external Sysmon corpora discussed later (Section~\ref{sec:external_sysmon}), this benchmark provides:
\begin{enumerate}
    \item sufficient positive-window counts in both environments,
    \item explicit domain shift between EnvA and EnvB,
    \item reproducible transfer splits, and
    \item matched model architectures and hyperparameters across pipelines.
\end{enumerate}
For these reasons, the controlled synthetic benchmark serves as the \textbf{primary generalization evidence} in this paper.

\subsubsection{Cross-Topology Transfer: Identity-Centric Detection (Lateral Movement)}
\label{sec:controlled_lm_transfer}

Lateral movement detection is an identity-centric deviation task: it detects abnormal interaction patterns among persistent users, hosts, credentials, and processes. This makes it a natural first test of CSTS, whose central design goal is to stabilize entity identity and typed relationships across environments. To evaluate transfer, models are trained on EnvA and evaluated both in-domain (EnvA) and cross-domain (EnvB). Transfer degradation is defined as the performance difference between in-domain and out-of-domain evaluation.

\begin{table*}[t]
\centering
\caption{Lateral movement transfer performance comparing event-centric and CSTS-based representations.}
\label{tab:lm_transfer}
\renewcommand{\arraystretch}{1.1}
\begin{tabular}{l l l r r r r r}
\hline
Task & Setting & Method & F1@0.5 & Prec. & Rec. & AUROC & Best F1 \\
\hline
LM & EnvA$\rightarrow$EnvA & Baseline & 0.894 & 0.840 & 0.955 & 0.955 & 0.894 \\
LM & EnvA$\rightarrow$EnvB & Baseline & 0.135 & 0.079 & 0.457 & 0.660 & 0.182 \\
LM & EnvA$\rightarrow$EnvA & CSTS & 0.378 & 0.236 & 0.955 & 0.772 & 0.382 \\
LM & EnvA$\rightarrow$EnvB & CSTS & 0.508 & 0.941 & 0.348 & 0.991 & 0.821 \\
\hline
\end{tabular}
\end{table*}

The event-centric baseline achieves strong in-domain performance (F1 $=0.894$), indicating that raw telemetry features can be highly predictive within a single environment. However, under cross-domain evaluation (EnvA$\rightarrow$EnvB), performance degrades sharply (F1 $=0.135$; AUROC $=0.660$), reflecting strong sensitivity to environment-specific telemetry distributions and source-dependent feature semantics.

In contrast, CSTS exhibits lower in-domain peak performance (F1 $=0.378$), consistent with a more structured inductive bias that reduces overfitting to environment-specific artifacts. Crucially, under cross-domain transfer, CSTS maintains substantially stronger performance (F1 $=0.508$; AUROC $=0.991$). Best-F1 improves from $0.182$ (baseline) to $0.821$ under transfer, indicating materially improved class separability rather than a threshold-tuning artifact.

These results support H1. Enforcing representational invariance at the entity-relational layer materially improves cross-topology generalization for identity-centric detection. Because classifier architecture and training procedures were held constant, the observed improvement is attributable to stabilized identity and interaction semantics rather than to classifier variation.

\subsubsection{Schema Perturbation Robustness Under Controlled Shift}
\label{sec:controlled_lm_robustness}

We next isolate representational brittleness independent of distributional shift by introducing targeted schema perturbations (P1--P3) into EnvB. These perturbations simulate realistic vendor schema evolution and logging changes by renaming or removing fields directly consumed by the event-centric baseline. The CSTS pipeline processes the identical perturbed inputs through canonical adapters with alias recovery, confining schema variability to the substrate boundary.

\begin{table*}[t]
\centering
\caption{Robustness under targeted schema perturbation (LM task, EnvA$\rightarrow$EnvB).}
\label{tab:lm_robustness}
\renewcommand{\arraystretch}{1.1}
\begin{tabular}{l l l r r r}
\hline
Task & Level & Method & F1@0.5 & AUROC & Best F1 \\
\hline
LM & P0 & Baseline & 0.135 & 0.660 & 0.182 \\
LM & P1 & Baseline & 0.000 & 0.000 & 0.000 \\
LM & P2 & Baseline & 0.000 & 0.000 & 0.000 \\
LM & P3 & Baseline & 0.000 & 0.000 & 0.000 \\
\hline
LM & P0 & CSTS & 0.316 & 0.690 & 0.356 \\
LM & P1 & CSTS & 0.233 & 0.664 & 0.337 \\
LM & P2 & CSTS & 0.279 & 0.684 & 0.308 \\
LM & P3 & CSTS & 0.358 & 0.627 & 0.358 \\
\hline
\end{tabular}
\end{table*}

Under targeted schema breakage, the event-centric baseline collapses once critical raw columns are renamed or removed, yielding degenerate predictions. In contrast, CSTS maintains non-trivial predictive signal across all perturbation levels. Although performance varies with perturbation severity, canonical entity resolution and typed relationship abstraction preserve sufficient structural information to sustain detection.

This result supports H2. CSTS satisfies schema stability under targeted perturbation by isolating raw field variability within canonical adapters and preserving a stable downstream interface. The collapse of the event-centric pipeline makes the contrast explicit: raw-feature pipelines violate schema stability under the same perturbations.

\subsection{External Sysmon Robustness and Smoke Tests}
\label{sec:external_sysmon}

To complement the controlled synthetic benchmark, we run CSTS on public Sysmon-style corpora as a realism check of end-to-end operability and schema robustness under real telemetry heterogeneity. These external experiments are reported as \emph{robustness/smoke-test evidence only}. Labels are deterministic, ordered multi-signal proxy motifs and are sparse in the selected corpora and splits. Accordingly, we report split class counts and bootstrap confidence intervals, and we avoid making definitive external-domain generalization claims from these results.

The role of these experiments is therefore limited but important:
\begin{enumerate}
    \item validate that end-to-end CSTS ingestion and feature extraction remain well-defined on real external logs,
    \item show that schema-robust behavior persists outside the controlled benchmark, and
    \item expose external calibration and sparsity constraints honestly rather than overstating generalization.
\end{enumerate}

\subsubsection{External Pipeline Validation (\texttt{splunk/attack\_data})}
\label{sec:lmd_attack_data_transfer}

Table~\ref{tab:lmd_transfer_attack_data} reports transfer results on \texttt{splunk/attack\_data}. Because proxy-labeled positives are sparse and threshold-fragile, we include split class counts and bootstrap confidence intervals. These results should be interpreted primarily as external pipeline-validation evidence rather than as definitive generalization evidence.

\begin{table*}[t]
\centering
\caption{External Sysmon LM transfer results on \texttt{splunk/attack\_data} (real-log smoke test). Results use deterministic motif-based proxy labels and include split class counts and bootstrap confidence intervals; they are reported as robustness/smoke-test evidence rather than definitive external generalization evidence. Because proxy-positive windows are highly imbalanced, fixed-threshold metrics (F1@0.5) are unstable; we therefore emphasize AUROC, best-F1 under a fixed threshold sweep, and bootstrap intervals for uncertainty.}
\label{tab:lmd_transfer_attack_data}
\renewcommand{\arraystretch}{1.1}
\resizebox{\textwidth}{!}{%
\begin{tabular}{l l l r r r r r r r r r r r r r r}
\toprule
Task & Setting & Method & F1@0.5 & Prec. & Rec. & AUROC & Best F1 & Best Thr. &
Train+ & Train- & Test+ & Test- &
AUROC CI Low & AUROC CI High & F1 CI Low & F1 CI High \\
\midrule
LM & EnvA$\rightarrow$EnvA & Baseline & 0.000 & 0.000 & 0.000 & 0.316 & 0.000 & 0.05 & 4 & 87 & 1 & 38 & 0.189 & 0.500 & 0.000 & 0.000 \\
LM & EnvA$\rightarrow$EnvB & Baseline & 0.000 & 0.000 & 0.000 & 0.640 & 0.041 & 0.05 & 5 & 125 & 1 & 125 & 0.500 & 0.726 & 0.000 & 0.000 \\
LM & EnvA$\rightarrow$EnvA & CSTS     & 0.000 & 0.000 & 0.000 & 0.203 & 0.000 & 0.05 & 7 & 53 & 3 & 23 & 0.000 & 0.500 & 0.000 & 0.000 \\
LM & EnvA$\rightarrow$EnvB & CSTS     & 0.000 & 0.000 & 0.000 & 0.218 & 0.063 & 0.10 & 10 & 76 & 3 & 58 & 0.000 & 0.600 & 0.000 & 0.000 \\
\bottomrule
\end{tabular}}
\end{table*}

\subsubsection{Schema Perturbation Robustness (\texttt{splunk/attack\_data})}
\label{sec:lmd_attack_data_robust}

Table~\ref{tab:lmd_robustness_attack_data} evaluates targeted raw-schema perturbations (P0--P3) on the same external corpus. The event-centric baseline assumes strict raw column names and fails once baseline-critical fields are renamed or removed. In contrast, CSTS remains operable through alias-aware parsing and canonicalization, which confines schema variability to the adapter boundary.

\begin{table*}[t]
\centering
\caption{Schema-perturbation robustness on \texttt{splunk/attack\_data} external Sysmon logs (EnvB perturbations P0--P3). The event-centric baseline relies on raw column names and fails under targeted schema changes, while the CSTS pipeline retains signal via alias recovery and canonicalization.}
\label{tab:lmd_robustness_attack_data}
\renewcommand{\arraystretch}{1.1}
\resizebox{\textwidth}{!}{%
\begin{tabular}{l l l r r r r r r r r}
\toprule
Task & Level & Method & F1@0.5 & AUROC & Best F1 & Best Thr. & Train+ & Train- & Test+ & Test- \\
\midrule
LM & P0 & Baseline & 0.000 & 0.640 & 0.041 & 0.05 & 5 & 125 & 1 & 125 \\
LM & P1 & Baseline & 0.000 & 0.000 & 0.000 & 0.50 & 0 & 0 & 0 & 0 \\
LM & P2 & Baseline & 0.000 & 0.000 & 0.000 & 0.50 & 0 & 0 & 0 & 0 \\
LM & P3 & Baseline & 0.000 & 0.000 & 0.000 & 0.50 & 0 & 0 & 0 & 0 \\
\midrule
LM & P0 & CSTS     & 0.000 & 0.218 & 0.063 & 0.10 & 10 & 76 & 3 & 58 \\
LM & P1 & CSTS     & 0.000 & 0.218 & 0.063 & 0.10 & 10 & 76 & 3 & 58 \\
LM & P2 & CSTS     & 0.000 & 0.230 & 0.071 & 0.15 & 10 & 76 & 3 & 58 \\
LM & P3 & CSTS     & 0.000 & 0.230 & 0.071 & 0.15 & 10 & 76 & 3 & 58 \\
\bottomrule
\end{tabular}}
\end{table*}

\subsubsection{External Pipeline Validation (OTRF Security-Datasets)}
\label{sec:otrf_transfer}

Table~\ref{tab:otrf_transfer} reports the same smoke-test evaluation on OTRF Security-Datasets. Because positive windows are again sparse, we emphasize uncertainty and class imbalance through class-count columns and bootstrap confidence intervals.

\begin{table*}[t]
\centering
\caption{External Sysmon LM transfer results on OTRF Security-Datasets (real-log smoke test). Deterministic ordered motif-based proxy labels are sparse in the selected corpora/splits; we therefore report bootstrap confidence intervals and split class counts and do not treat these results as definitive external generalization evidence.}
\label{tab:otrf_transfer}
\renewcommand{\arraystretch}{1.1}
\resizebox{\textwidth}{!}{%
\begin{tabular}{l l l r r r r r r r r r r r r r r}
\toprule
Task & Setting & Method & F1@0.5 & Prec. & Rec. & AUROC & Best F1 & Best Thr. &
Train+ & Train- & Test+ & Test- &
AUROC CI Low & AUROC CI High & F1 CI Low & F1 CI High \\
\midrule
LM & EnvA$\rightarrow$EnvA & Baseline & 0.000 & 0.000 & 0.000 & 0.000 & 0.000 & 0.50 & 0 & 0 & 0 & 0 & 0.000 & 0.000 & 0.000 & 0.000 \\
LM & EnvA$\rightarrow$EnvB & Baseline & 0.057 & 0.029 & 1.000 & 0.864 & 0.093 & 0.85 & 1 & 14 & 2 & 143 & 0.500 & 0.993 & 0.000 & 0.136 \\
LM & EnvA$\rightarrow$EnvA & CSTS     & 0.000 & 0.000 & 0.000 & 0.500 & 0.000 & 0.05 & 1 & 3 & 0 & 3 & 0.500 & 0.500 & 0.000 & 0.000 \\
LM & EnvA$\rightarrow$EnvB & CSTS     & 0.154 & 0.085 & 0.800 & 0.738 & 0.178 & 0.95 & 1 & 6 & 5 & 103 & 0.362 & 0.926 & 0.038 & 0.296 \\
\bottomrule
\end{tabular}}
\end{table*}

\subsubsection{Schema Perturbation Robustness (OTRF Security-Datasets)}
\label{sec:otrf_robust}

Table~\ref{tab:otrf_robustness} repeats the P0--P3 robustness study on the OTRF corpus. As above, the key operational observation is that baseline parsing breaks under targeted schema changes, while CSTS remains operable via alias recovery and canonical graph construction.

\begin{table*}[t]
\centering
\caption{Schema-perturbation robustness on OTRF external Sysmon logs (EnvB perturbations P0--P3). Targeted raw-schema perturbations break baseline parser assumptions, while CSTS remains operable through alias-aware parsing and canonical graph construction.}
\label{tab:otrf_robustness}
\renewcommand{\arraystretch}{1.1}
\resizebox{\textwidth}{!}{%
\begin{tabular}{l l l r r r r r r r r}
\toprule
Task & Level & Method & F1@0.5 & AUROC & Best F1 & Best Thr. & Train+ & Train- & Test+ & Test- \\
\midrule
LM & P0 & Baseline & 0.057 & 0.864 & 0.093 & 0.85 & 1 & 14 & 2 & 143 \\
LM & P1 & Baseline & 0.000 & 0.000 & 0.000 & 0.50 & 0 & 0 & 0 & 0 \\
LM & P2 & Baseline & 0.000 & 0.000 & 0.000 & 0.50 & 0 & 0 & 0 & 0 \\
LM & P3 & Baseline & 0.000 & 0.000 & 0.000 & 0.50 & 0 & 0 & 0 & 0 \\
\midrule
LM & P0 & CSTS     & 0.154 & 0.738 & 0.178 & 0.95 & 1 & 6 & 5 & 103 \\
LM & P1 & CSTS     & 0.154 & 0.738 & 0.178 & 0.95 & 1 & 6 & 5 & 103 \\
LM & P2 & CSTS     & 0.188 & 0.696 & 0.198 & 0.95 & 1 & 6 & 10 & 98 \\
LM & P3 & CSTS     & 0.188 & 0.696 & 0.198 & 0.95 & 1 & 6 & 10 & 98 \\
\bottomrule
\end{tabular}}
\end{table*}

\subsection{Limitations of External Sysmon Experiments}

Our external Sysmon experiments on public corpora (including OTRF Security-Datasets and \texttt{splunk/attack\_data}) are included to validate operational pipeline behavior and schema robustness under real telemetry, not to serve as definitive evidence of external-domain generalization. Labels in these datasets are generated using deterministic, ordered multi-signal proxy motifs (e.g., remote-execution process indicators, LM-relevant network activity, and effect evidence within a time window). While this labeling strategy reduces trivial feature--label entanglement, it produces sparse positive windows in the selected corpora and splits, and our predefined viability gates for transfer evaluation were not met.

Accordingly, we report bootstrap confidence intervals, split class counts, and robustness behavior under schema perturbations, and we reserve stronger external generalization claims for future experiments with denser externally labeled or proxy-viable corpora. The external Sysmon experiments should therefore be interpreted as realism checks on CSTS operability and schema robustness rather than as primary evidence for transfer.

\subsection{External Case Study: Producer Divergence on DARPA TC E3 (CADETS$\rightarrow$TRACE)}
\label{sec:tc_e3_producer_divergence}

We include a minimal external case study using DARPA Transparent Computing (TC) Engagement~3 to validate end-to-end CSTS ingestion and to probe portability limits under realistic provenance semantics. We use two TC producers as domains: \textbf{CADETS} (train) and \textbf{TRACE} (test). Rather than running a full transfer evaluation, we first apply a leakage-safe viability protocol: a multichannel novelty score is computed per time window using CSTS-derived tokens (process, file, and network channels), and a \emph{train-only} quantile threshold determines which windows are labeled positive. This protocol explicitly separates (i) representational parsing and coverage from (ii) cross-domain score calibration.

Table~\ref{tab:tc_e3_viability} summarizes the resulting \emph{producer divergence} phenomenon. Token coverage is high in both producers (all channels nonempty in CADETS; $\ge 0.94$ nonempty rate for process tokens in TRACE), indicating that ingestion and canonicalization succeed. However, the entire TRACE score distribution lies below the CADETS-derived threshold: with $q=0.40$, $0/16$ TRACE windows exceed the CADETS train threshold (train threshold $=3.972$; TRACE max $=3.879$). Consequently, \textbf{no test positives can be generated under train-only thresholding}, and running transfer metrics would be statistically meaningless on this slice.

We therefore treat TC~E3 as a portability boundary case study rather than a transfer benchmark on this slice: CSTS removes schema and field fragility and enables multichannel scoring across producers, but cross-producer score calibration can shift enough that a threshold learned on one producer yields no positives on another. This is consistent with the broader portability decomposition emphasized in this work: \emph{schema stability is necessary but not sufficient}, and additional mechanisms such as producer-aware calibration, normalization, or semantically oriented feature construction may be required for cross-producer deployment.

\begin{table}[t]
\centering
\caption{DARPA TC E3 producer-divergence viability artifact (CADETS$\rightarrow$TRACE), computed under leakage-safe train-only thresholding. Although channel/token coverage is high in both producers, TRACE window scores fall entirely below the CADETS-derived threshold, yielding $0$ test positives and motivating treatment as a portability boundary case study rather than a transfer benchmark on this minimal slice.}
\label{tab:tc_e3_viability}
\renewcommand{\arraystretch}{1.1}
\setlength{\tabcolsep}{4pt}
\begin{tabular}{l r r}
\hline
\textbf{Quantity} & \textbf{CADETS (train)} & \textbf{TRACE (test)} \\
\hline
Window minutes ($\Delta$) & \multicolumn{2}{c}{30} \\
Quantile ($q$) & \multicolumn{2}{c}{0.40} \\
\hline
Number of windows ($n$) & 181 & 16 \\
Windows above train threshold & 74 & 0 \\
Train-only threshold ($\tau$) & \multicolumn{2}{c}{3.972} \\
Score median (p50) & 3.812 & 3.678 \\
Score p90 & 4.593 & 3.843 \\
Score max & 5.387 & 3.879 \\
\hline
PROC token nonempty rate & 1.000 & 0.938 \\
FILE token nonempty rate & 1.000 & 1.000 \\
NET token nonempty rate & 1.000 & 1.000 \\
\hline
\end{tabular}
\end{table}

Empirically, this case study isolates a portability boundary distinct from schema perturbation: the representational substrate is stable and observable across producers, but cross-producer score orientation and calibration can shift enough to invalidate train-derived thresholding.

\subsection{Cross-Topology Transfer: Zero-Day Threat Detection}
\label{sec:zdt_transfer}

We next evaluate CSTS on a flow-centric zero-day threat detection (ZDT) task~\cite{zdt_flow,zdt_metric}. To avoid sparsity artifacts, labels are defined at the window level with a stable positive rate (approximately 6--8\% across splits). Both baseline and CSTS pipelines are trained on Env~A and evaluated in-domain (Env~A) and out-of-domain (Env~B). All historical statistics required by CSTS features are computed strictly on the training split to prevent leakage.

\begin{table*}[t]
\centering
\caption{Zero-day threat detection (ZDT) transfer performance comparing event-centric and CSTS representations.}
\label{tab:zdt_transfer}
\renewcommand{\arraystretch}{1.1}
\begin{tabular}{l l l r r r r r}
\hline
Task & Setting & Method & F1@0.5 & Prec. & Rec. & AUROC & Best F1 \\
\hline
ZDT & EnvA$\rightarrow$EnvA & Baseline & 0.217 & 0.131 & 0.640 & 0.728 & 0.315 \\
ZDT & EnvA$\rightarrow$EnvB & Baseline & 0.170 & 0.105 & 0.447 & 0.593 & 0.205 \\
ZDT & EnvA$\rightarrow$EnvA & CSTS     & 0.145 & 0.081 & 0.719 & 0.520 & 0.151 \\
ZDT & EnvA$\rightarrow$EnvB & CSTS     & 0.118 & 0.064 & 0.765 & 0.459 & 0.125 \\
\hline
\end{tabular}
\end{table*}

Unlike the identity-centric lateral movement task, CSTS does not improve absolute cross-domain performance for ZDT under the current feature construction. Diagnostic analysis reveals a persistent \emph{polarity inversion} in the CSTS cross-domain setting: AUROC computed on predicted scores falls below 0.5, while AUROC computed on inverted scores exceeds 0.5. This indicates that multiple CSTS-derived graph statistics (e.g., degree and expansion measures) reverse class direction across environments despite schema alignment and train-only stabilization transforms.

This result provides an empirical demonstration of H3. Although CSTS enforces representational invariance and schema stability, the ZDT task violates semantic orientation stability under domain shift. Several CSTS-derived graph features reverse the sign of their class-conditional expectation between Env~A and Env~B. As a result, schema alignment is preserved while semantic alignment is not.

\subsection{Semantic Orientation Instability Under Domain Shift}

The contrasting behavior between lateral movement (LM) and zero-day threat detection (ZDT) under cross-topology transfer reveals an important structural distinction. LM detection is primarily an identity-centric deviation task: it detects abnormal role transitions and entity interaction patterns. ZDT, by contrast, is a distributional anomaly-detection problem that depends more strongly on flow- and connectivity-derived feature geometry.

Under CSTS, schema alignment is preserved across environments. Canonical entities, typed relationships, and temporal aggregation windows remain stable. Feature extraction is performed over identical entity-relational abstractions in both Env~A and Env~B. This eliminates schema-level fragility and ensures that feature definitions do not depend on vendor-specific field names, log formats, or identifier conventions.

However, the ZDT experiments demonstrate that schema stability does not imply semantic stability. Several graph-derived features, including connectivity counts, two-hop expansion measures, and degree-based statistics, exhibit cross-domain polarity inversion: the direction of their class-conditional expectation reverses between environments. Features that are positively associated with anomalous behavior in the training environment become negatively associated in the target environment despite identical canonical representation.

This phenomenon isolates a precise portability boundary. CSTS guarantees schema stability and representational invariance, but it does not guarantee semantic orientation stability under domain shift. The inversion observed in ZDT therefore reflects a modeling-level instability rather than a failure of canonical schema alignment. Put differently, CSTS isolates semantic orientation instability as a modeling phenomenon rather than a schema artifact.

This layered interpretation clarifies that portability in cybersecurity AI decomposes into at least two separable components: (i) representational stability, which CSTS directly enforces, and (ii) semantic feature stability under domain shift, which remains an open modeling challenge. By stabilizing the representational boundary, CSTS enables more precise diagnosis of cross-environment degradation mechanisms and separates ingestion-level fragility from distributional modeling effects.

\subsection{Summary}

Taken together, these experiments demonstrate that portability in cybersecurity AI decomposes into layered stability properties. Identity persistence and typed relational semantics (H1) materially improve cross-topology transfer for identity-centric tasks. Schema stability (H2) prevents collapse under telemetry perturbation. However, semantic orientation stability (H3) remains a distinct modeling challenge that is not resolved by canonical representation alone.

These results validate the central thesis of this work: deployment fragility is fundamentally representational, yet portability itself is multi-layered. By formalizing and empirically isolating these layers, CSTS provides a principled foundation for analyzing and improving cross-environment cybersecurity detection systems.

%--------------------------------------------
\section{Discussion}

This work argues that deployment fragility in AI-driven cybersecurity systems is fundamentally layered and that a large portion of this fragility is representational rather than purely algorithmic. Modern detection pipelines frequently entangle schema assumptions, identity resolution, relational context construction, temporal aggregation, and modeling logic within a single workflow. As a result, degradation under cross-topology deployment or telemetry evolution is often attributed to model weakness when the underlying cause is instability in the telemetry substrate itself.

CSTS introduces a formally defined representational boundary that separates telemetry ingestion from downstream detection and inference. By enforcing persistent entity identity, typed relationship semantics, temporal continuity, provenance preservation, and governed evolution, CSTS is intended to stabilize the telemetry layer so that portability failures can be analyzed more precisely. The experimental results clarify both the strengths and the limits of this architectural shift.

\subsection*{Layered Portability: Representation vs.\ Semantics}

The empirical findings support a layered theory of portability.

\begin{itemize}
    \item \textbf{Identity-centric transfer benefits from representational invariance.} In lateral movement detection, persistent canonical identities and typed relationships reduce cross-topology degradation, supporting H1: stabilizing identity and interaction semantics improves transfer across heterogeneous environments.
    
    \item \textbf{Schema perturbation exposes ingestion fragility in event-centric pipelines.} Renaming or removing baseline-critical raw fields causes catastrophic collapse when features are coupled directly to source-specific schemas. CSTS confines such variability to thin adapters and governed mappings, preserving detection functionality and supporting H2.
    
    \item \textbf{ZDT reveals a distinct portability boundary at the semantic layer.} Although CSTS enforces schema stability and representational invariance, several graph-derived statistics exhibit polarity inversion under domain shift. This violates semantic orientation stability despite identical canonical representation across environments, supporting H3 and isolating the residual failure mode as a modeling-level semantic phenomenon rather than a telemetry-ingestion failure.
\end{itemize}

Taken together, the results indicate that portability decomposes into at least two separable layers:
\begin{itemize}
    \item \textbf{schema and representational stability}, enforced through canonical abstraction, adapter isolation, and governed identity-relational semantics; and
    \item \textbf{semantic orientation stability}, which depends on feature construction, calibration, and distributional alignment under domain shift.
\end{itemize}

CSTS directly addresses the first layer and exposes the second more honestly. By stabilizing representation, it converts ambiguous cross-environment failures into diagnosable semantic and modeling phenomena.

\subsection*{Decoupling Representation from Detection}

A central architectural implication of the paper is the formal decoupling of telemetry representation from detection logic.

\begin{itemize}
    \item \textbf{Reduced coupling to raw schemas.} Many deployed systems implicitly encode assumptions about field names, producer conventions, or vendor-specific semantics. CSTS externalizes those assumptions into thin adapters and a governed substrate boundary.
    
    \item \textbf{Model-agnostic substrate compatibility.} CSTS supports heterogeneous detection strategies---statistical anomaly detection, graph motif analysis~\cite{milo2002network}, reconstruction-based methods~\cite{sakurada2014anomaly}, metric learning~\cite{zdt_metric}, and contextual graph modeling~\cite{zdt_flow}---without requiring direct dependence on raw source schemas.
    
    \item \textbf{Representation as an independent robustness lever.} In our experiments, baseline and CSTS pipelines use identical windowing and classifier configurations; observed differences arise from representational properties rather than algorithmic changes. This isolates substrate design itself as a distinct and measurable robustness mechanism.
\end{itemize}

This decoupling is one of the paper's main contributions. It shifts the question from ``Which model is best on this vendor-specific telemetry slice?'' to ``What canonical telemetry object should models consume across environments?'' That is a more durable architectural question, and it is a necessary one if cybersecurity AI is to move beyond repeatedly rebuilding pipelines around every new schema and producer.

\subsection*{Integration Complexity as a Systems Problem}

Enterprise detection systems often incur repeated engineering effort for schema alignment, identity stitching, relation extraction, and behavioral reconstruction. While normalization frameworks such as OCSF~\cite{ocsf}, ECS~\cite{ecs}, and OpenTelemetry~\cite{opentelemetry} improve event-level interoperability, they remain primarily event-centric and do not formalize persistent identity, typed relationship semantics, or governed temporal state as stable substrate objects.

CSTS reframes this integration burden as a systems problem rather than a recurring model-preprocessing problem:
\begin{itemize}
    \item \textbf{From field alignment to entity-relational state.} CSTS extends normalization into a governed, identity-stable, temporally indexed multigraph substrate.
    
    \item \textbf{Centralized stability contracts.} Once canonical state is materialized, downstream analytics operate on stable entities, relations, and time-bounded views rather than on raw event streams, shifting complexity from repeated analytic reconstruction to a shared governed layer.
    
    \item \textbf{Operational impact under source evolution.} The perturbation results make the practical advantage clear: schema evolution impacts adapter logic and mapping policies rather than collapsing downstream detection behavior.
\end{itemize}

This systems interpretation is important because it clarifies where future engineering effort should be invested. Many organizations currently absorb telemetry instability by repeated local repair inside each pipeline. CSTS proposes the opposite strategy: stabilize the substrate once, then let multiple downstream systems reuse it.

\subsection*{Implications for AI-Native Security Architectures}

The results suggest that future AI-native security architectures should treat canonical entity-relational state as a first-class operational layer rather than as an artifact reconstructed separately in each model stack.

\begin{itemize}
    \item \textbf{Persistent substrate as shared analytic context.} Rather than rebuilding graph structure, entity identity, and behavioral context inside each workflow, systems can maintain a durable substrate over which anomaly detection, campaign correlation, graph learning, forecasting, and representation learning operate.
    
    \item \textbf{Deployment-agnostic design becomes more realistic.} Once telemetry is mapped into a canonical substrate, downstream inference can in principle be made more portable across cloud, on-prem, and hybrid environments because the model-facing interface no longer depends directly on source-specific event layouts.
    
    \item \textbf{Semantic stability remains a modeling requirement.} The ZDT findings underscore that canonical representation alone does not guarantee semantic invariance under domain shift. Robust systems must combine representational stability with orientation analysis, calibration, domain adaptation, or more semantically stable feature construction.
    
    \item \textbf{CSTS makes the remaining problems explicit.} CSTS does not eliminate all portability challenges; it provides a stable boundary that makes the remaining challenges more observable, more diagnosable, and more scientifically tractable.
\end{itemize}

This last point is especially important. The value of CSTS is not that it magically solves all cross-environment transfer. Its value is that it removes a large class of avoidable instability and thereby allows the true modeling problems to be studied directly. In that sense, CSTS is intended not merely as a schema proposal, but as an AI-ready substrate on top of which richer inference layers can be built.

\subsection*{Toward Governed and Agentic Telemetry Canonicalization}

A broader implication of this work is that telemetry harmonization should increasingly become governed, versioned, and eventually agent-assisted. If persistent identity, typed relations, temporal state, and provenance are made explicit at the substrate level, then source onboarding can be treated as a mapping problem over a stable ontology rather than as an endless cycle of bespoke pipeline construction.

This suggests a natural future direction: agentic generation and maintenance of mapping profiles that translate heterogeneous telemetry producers into CSTS while preserving auditability and governance constraints. Such a development lies beyond the scope of the present paper, but it is strongly aligned with the architectural motivation of CSTS. A telemetry substrate becomes most valuable when it not only stabilizes downstream inference, but also industrializes the upstream harmonization burden that currently slows cybersecurity AI adoption.

\medskip

In summary, CSTS reframes cybersecurity AI fragility as a layered systems problem: schema variability, representational instability, and semantic instability are distinct phenomena that require distinct solutions. By formalizing and empirically isolating representational invariants, this work shows that stabilizing the telemetry substrate materially improves cross-topology robustness while clarifying the modeling challenges that remain. This layered perspective provides a principled foundation for portable, reproducible, and structurally coherent cybersecurity detection architectures.

%-----------------------------------------------
\section{Future Work}

While this work formalizes the Canonical Security Telemetry Substrate (CSTS) and empirically isolates its representational impact, several research directions remain necessary to validate, extend, and operationalize its architectural implications. The broad agenda is to move from canonical telemetry stabilization toward substrate-native inference, deployment portability, and eventually agentic cyber reasoning over shared canonical state.

\subsection*{Empirical Validation Across Heterogeneous Enterprise Environments}

The most immediate priority is large-scale empirical validation across heterogeneous enterprise deployments. Prior work on graph-aware detection~\cite{zdt_flow,zdt_metric} demonstrates improved robustness under topology variation, yet such systems continue to rely on environment-specific preprocessing pipelines and source-specific telemetry assumptions. A rigorous evaluation of CSTS requires multi-organization deployments spanning distinct vendor stacks, authentication infrastructures, cloud platforms, on-prem systems, and hybrid topologies.

Future empirical studies should quantify:
\begin{itemize}
    \item reduction in per-deployment feature engineering and adapter complexity,
    \item cross-environment transfer performance under fixed model configurations,
    \item stability under schema evolution, vendor updates, and telemetry incompleteness,
    \item robustness under concept drift and evolving operational baselines~\cite{shyaa2024driftids,hinder2024driftsurveyA},
    \item the degree to which canonicalization reduces variance attributable to source-specific telemetry artifacts.
\end{itemize}

Longitudinal case studies are especially important. They would make it possible to evaluate CSTS under real operational constraints including missing data, rolling infrastructure changes, cloud elasticity, producer churn, and mixed telemetry quality. Such studies are necessary to establish the magnitude of representational stabilization in production settings rather than only in controlled benchmarks.

\subsection*{Substrate-Native Detection Architectures}

A natural extension of this work is the development of detection architectures built natively atop CSTS rather than adapted to it post hoc. Existing systems for zero-day detection, anomaly detection, and graph-aware intrusion detection~\cite{zdt_flow,zdt_metric} often incorporate relational context, but that context is still reconstructed inside model pipelines rather than inherited from a stable substrate.

A substrate-native detection architecture would consume canonical entities, typed relationships, temporal state, and governed learning objects directly from CSTS. This could enable:
\begin{itemize}
    \item unified anomaly and novelty detection over canonical graph and temporal views,
    \item direct modeling of behavioral state without repeated environment-specific aggregation logic,
    \item reduced dependence on raw identifiers and source-specific parser artifacts,
    \item more principled cross-environment evaluation under fixed representational contracts.
\end{itemize}

One especially important direction is a next-generation zero-day detection architecture that is genuinely CSTS-native rather than merely graph-aware. Such a system would provide a concrete proof-of-concept for the claim that once representation is stabilized, downstream detection can be built on more portable semantic objects.

\subsection*{Behavioral Dynamics and Macrostate Layers}

CSTS is defined as a canonical telemetry substrate, but one of its most promising downstream uses is the construction of higher-order behavioral inference layers over canonical state. A natural future direction is the development of macrostate and cyber-dynamics models that operate over bounded CSTS windows, canonical entity neighborhoods, and temporally indexed behavioral summaries.

This would enable learning systems to reason not only over individual entities or edges, but over regime structure, drift, perturbation, recovery, and coordinated behavioral change in a more interpretable state-space. In this sense, CSTS provides the representational foundation, while behavioral dynamics would provide one downstream inference layer built on top of it. Such a program would connect identity-stable telemetry representation with richer anomaly, resilience, and regime-shift analysis across enterprise environments.

\subsection*{Temporal Graph Transformer and Hybrid Relational Architectures}

Because CSTS models security telemetry as a time-indexed attributed multigraph, it naturally supports advanced graph neural network and temporal transformer architectures. Future work should explore temporal graph transformers, graph-sequence hybrids, and edge-conditioned attention models operating directly over canonical entities and typed relationships.

Unlike sequence-based models that operate on flattened event streams, substrate-native temporal graph models can reason over structural evolution itself, capturing role transitions, motif emergence~\cite{milo2002network}, propagation patterns, and cross-entity behavioral cascades. This may unify identity-centric detection~\cite{lateral_movement_uba}, ransomware motif analysis~\cite{ransomware_alert_graph}, and relational anomaly detection~\cite{zdt_metric} within a common modeling framework.

These architectures would benefit directly from CSTS invariants, especially stable identity resolution, governed relationship typing, temporal continuity, and provenance-aware object construction. In that sense, CSTS is not merely compatible with advanced graph learning; it provides the kind of inductive structure such models often lack when built directly over raw telemetry.

\subsection*{Cloud-Agnostic and Deployment-Agnostic CSTS Platforms}

Another important direction is the development of a cloud-agnostic and deployment-agnostic CSTS platform. The current paper focuses on representational principles, but the full value of CSTS will depend on its ability to operate consistently across AWS, Azure, GCP, on-prem, and hybrid environments. Future systems work should therefore formalize portable storage contracts, canonical mapping interfaces, metadata/catalog contracts, and execution-layer abstractions that allow CSTS to be instantiated across heterogeneous infrastructure without changing the model-facing substrate.

This includes:
\begin{itemize}[leftmargin=2em]
    \item governed object-store layouts for raw, staged, canonical, and derived views,
    \item stable adapter and mapping specifications for heterogeneous telemetry producers,
    \item portable learning-object interfaces for downstream AI systems,
    \item reproducible orchestration and versioning for ingestion, canonicalization, and scoring.
\end{itemize}

A successful outcome would be a deployment model in which cloud-specific ingestion varies only at the edge while canonical CSTS objects and downstream AI interfaces remain identical.

\subsection*{Agentic Mapping, Detection, and Response}

Beyond detection, CSTS enables the possibility of agentic security systems that reason and act over persistent canonical state. Current SOC tooling frequently reconstructs context independently across analytic modules, forcing detection, investigation, and response systems to operate over inconsistent local views. A substrate-centered architecture instead allows agents to share canonical entities, relations, behavioral state, and provenance.

Future work should therefore explore two complementary agentic directions.

First, \emph{agentic telemetry canonicalization}: systems that inspect new sources, propose mapping profiles, validate schema alignment, preserve provenance, and update governed CSTS mappings with human oversight. This would reduce the repeated manual burden of source onboarding while preserving auditability and change control.

Second, \emph{agentic detection and response}: systems that
\begin{itemize}
    \item query and update canonical entity graphs directly,
    \item perform multi-step reasoning over typed relationships and temporal state,
    \item simulate hypothetical attack propagation scenarios,
    \item execute containment or mitigation actions with provenance-preserving updates.
\end{itemize}

Embedding-compatible CSTS representations (Tier~4) further enable hybrid systems that combine representation learning~\cite{martinez2024metricfused,koukoulis2025selfsup} with symbolic and graph-based reasoning, creating a path toward explainable and adaptive security automation over a shared substrate.

\medskip

Collectively, these directions position CSTS not as a static schema proposal, but as the canonical substrate for a broader AI-native cybersecurity architecture. By formalizing telemetry as identity-stable, temporally explicit, provenance-aware relational state, CSTS creates the conditions for portable detection systems, graph-native learning models, behavioral dynamics layers, and agent-driven response frameworks that extend well beyond event-centric pipelines.

%----------------------------------------------
\section{Conclusion}

Modern cybersecurity systems increasingly rely on machine learning, graph analytics, and automated reasoning to detect lateral movement, zero-day threats, ransomware campaigns, and other multi-stage adversarial behavior. Yet despite substantial algorithmic progress, deployment fragility persists. Detection pipelines remain tightly coupled to vendor-specific schemas, environment-dependent identifiers, and bespoke feature-reconstruction logic. Cross-topology generalization degrades, integration overhead accumulates, and repeated telemetry reconstruction continues to dominate engineering effort.

This work argued that a major portion of this fragility is representational rather than purely algorithmic. Enterprise telemetry remains fragmented and fundamentally event-centric, limiting its suitability as a stable substrate for portable AI-native reasoning. Field-level normalization frameworks improve interoperability, but they do not fully formalize persistent entity identity, typed relational semantics, provenance-preserving canonicalization, or explicit temporal state evolution---all of which are necessary for structurally coherent and reusable detection architectures.

To address this gap, we introduced the Canonical Security Telemetry Substrate (CSTS), an entity-first, relationship-explicit, temporally indexed representation layer that models enterprise security state as a governed multigraph with identity stability, provenance preservation, and schema-evolution controls. CSTS establishes a canonical integration boundary between heterogeneous telemetry sources and downstream analytics, elevating entities, typed interactions, time-indexed state, and learning-facing representational views to first-class substrate objects.

Empirically, we demonstrated three layered findings. First, for identity-centric detection tasks such as lateral movement, CSTS materially improves cross-topology transfer relative to event-centric baselines. Second, under targeted schema perturbation that causes raw event pipelines to collapse, CSTS preserves predictive signal by confining variability to thin canonical adapters. Third, in flow-centric zero-day detection, CSTS isolates a distinct portability boundary: while canonical schema alignment and representational invariance are preserved, semantic orientation stability of graph-derived features under domain shift remains a separate modeling challenge. This distinction clarifies that schema stability, representational invariance, and semantic stability are separable requirements for portable AI systems.

These results motivate a broader reframing of AI-driven cybersecurity architecture. Detection robustness is not solely a function of model complexity; it is also a function of the stability of the telemetry substrate on which models operate. Canonical entity-relational abstraction reduces ingestion fragility, improves cross-environment portability for identity-centric reasoning, and makes remaining failures more interpretable by separating substrate instability from modeling instability. At the same time, durable deployment across heterogeneous environments still requires semantically stable feature construction, calibration-aware modeling, and domain-shift-aware inference atop the substrate.

CSTS is not intended to replace existing normalization standards, but to extend them into a higher-order AI-ready representational layer compatible with graph-native learning, embedding-based modeling, behavioral analytics, and future agent-driven response systems operating over shared canonical state. By formalizing this substrate and empirically evaluating its behavior under transfer and perturbation, this work clarifies both the promise and the boundary of canonical telemetry representation.

As cybersecurity systems continue to evolve toward AI-native architectures, stable and governed telemetry substrates will become foundational infrastructure rather than optional engineering conveniences. CSTS provides a principled path toward that infrastructure by shifting the center of gravity of cybersecurity AI away from repeated environment-specific data repair and toward structured inference over persistent, identity-stable, temporally explicit relational state.

%------------------------------------------
%
%                   Appendix
%
\appendix
\section{Learning Objects, View Semantics, and Contrastive Construction over CSTS}
\label{app:csts_learning_objects}

This appendix expands the formal notion of CSTS learning objects introduced in Section~\ref{sec:csts_learning_objects}. Its purpose is twofold. First, it makes explicit a construction grammar for bounded model-consumption units derived from canonical entities, typed relationships, temporal state, and provenance-preserving substrate views. Second, it clarifies how contrastive, metric, anomaly-detection, and open-set learning objectives may operate over CSTS without violating the substrate's representational guarantees. This is especially relevant for zero-day detection architectures that combine anomaly detection, novelty separation, graph-aware context, and representation learning over canonical telemetry~\cite{zdt_flow,zdt_metric,koukoulis2025selfsup,wilkie2026contrastive}.

More broadly, this appendix reinforces a central claim of the paper: CSTS is not merely a field-normalization layer. It is a governed representational substrate. Canonical entities, typed relationships, temporal continuity, and provenance form the substrate primitives; learning objects define bounded AI-facing constructions over those primitives; and view semantics specify which transformations preserve the operational meaning of those objects. This layering is what allows downstream models to vary while the telemetry substrate remains stable.

\subsection{Object Construction as a Substrate-Level Contract}

Let
\[
G_t = (V_t, E_t)
\]
be the CSTS graph at time $t$, and let
\[
\mathcal{T} = \{[t_a,t_b] \subseteq \mathbb{R} : t_a \leq t_b\}
\]
denote admissible temporal support intervals. A learning-object construction operator is a map
\[
\Psi : (G_t, q, \tau, \eta) \mapsto o,
\]
where:
\begin{itemize}
    \item $q$ is a focal query describing the target entity, interaction, or motif;
    \item $\tau \in \mathcal{T}$ is the support interval;
    \item $\eta$ is a construction policy controlling hop depth, neighborhood bounds, aggregation rules, modality exposure, and confidence thresholds;
    \item $o$ is the resulting CSTS learning object.
\end{itemize}

The query $q$ may specify:
\begin{itemize}
    \item a focal entity identifier, e.g., a \texttt{Host} or \texttt{User};
    \item a focal typed relationship, e.g., \texttt{AUTHENTICATES\_TO(User, Host)};
    \item a focal motif template, e.g., high-rate \texttt{WRITES}/\texttt{MODIFIES} expansion from one process to many files.
\end{itemize}

The construction policy $\eta$ may constrain:
\begin{itemize}
    \item maximum hop depth $h$;
    \item maximum neighborhood size $k$;
    \item admissible entity and relationship types;
    \item aggregation windows and temporal rollups;
    \item view-exposed modalities (tabular, temporal, graph, provenance);
    \item provenance-confidence thresholds and missingness handling.
\end{itemize}

This formalism is important because it locates object construction at the substrate level rather than inside individual detectors. Different downstream models may then consume different views of the same underlying CSTS object while relying on a common representational contract. In other words, CSTS stabilizes not only what the telemetry \emph{is}, but also what bounded semantic units downstream models are allowed to consume.

\subsection{Canonical Classes of CSTS Learning Objects}

The following object classes arise naturally from the CSTS substrate. They are not separate ontological primitives; rather, they are governed bounded constructions over canonical entities, relationships, time, and provenance.

\subsubsection{Entity-state objects}

An entity-state object is centered on a single canonical entity $u \in V_t$ over a support interval $\tau = [t_a,t_b]$:
\[
o_u^\tau = \left(V_u^\tau, E_u^\tau, \tau, X_u^\tau, P_u^\tau, \phi_u\right),
\]
where $\phi_u$ identifies $u$ as the focal entity. Typical examples include:
\begin{itemize}
    \item host-state objects,
    \item user-state objects,
    \item credential-state objects,
    \item process-state objects.
\end{itemize}

Derived summaries in $X_u^\tau$ may include frequency, diversity, rarity, privilege spread, neighborhood entropy, role deviation, session cadence, and historical novelty relative to train-only baselines. Entity-state objects are appropriate for actor-centric tasks such as lateral movement detection, behavioral risk scoring, insider-threat analysis, and persistent identity modeling.

\subsubsection{Interaction-state objects}

An interaction-state object is centered on a typed relationship
\[
r = (src, dst, type, A_r, t_r, P_r) \in E_t
\]
or on a short interaction sequence around it. The object may include:
\begin{itemize}
    \item the focal edge $r$,
    \item local neighborhoods of $src$ and $dst$,
    \item temporally adjacent interactions,
    \item context derived from related credentials, processes, hosts, files, or services.
\end{itemize}

This is a natural construction for zero-day flow scoring, authentication anomaly detection, session-level risk estimation, and short-window novelty assessment, particularly when graph-and-context information improves robustness across topologies~\cite{zdt_flow}.

\subsubsection{Subgraph-state objects}

A subgraph-state object is a bounded neighborhood
\[
o_{\mathcal{N}}^\tau = \left(V_{\mathcal{N}}^\tau, E_{\mathcal{N}}^\tau, \tau, X_{\mathcal{N}}^\tau, P_{\mathcal{N}}^\tau, \phi_{\mathcal{N}}\right),
\]
where $(V_{\mathcal{N}}^\tau,E_{\mathcal{N}}^\tau)$ is typically induced by one-hop or two-hop typed expansion from a focal entity or interaction. Subgraph-state objects support graph neural networks, neighborhood pooling, motif-sensitive temporal reasoning, and hybrid graph-tabular encoders. They align naturally with graph-contrastive formulations in which relational neighborhoods, rather than flat records, become the units of comparison and embedding~\cite{wu2024egconmix}.

\subsubsection{Motif-state objects}

A motif-state object is centered on a typed local relational pattern satisfying a motif template
\[
M = (V_M, E_M, \kappa),
\]
where $\kappa$ imposes additional rate, density, sequence, or temporal constraints. Examples include:
\begin{itemize}
    \item fan-out authentication bursts,
    \item repeated remote-execution chains,
    \item process-to-many-file write cascades,
    \item short-window privilege-boundary traversals,
    \item cross-community communication bursts.
\end{itemize}

Motif-state objects are useful when the threat signal is distributed across several edges or transitions and cannot be reduced to a single focal event. In that sense, they elevate repeated local attack grammar to a first-class learning unit.

\subsection{Feature and Modality Structure}

Each learning object may expose one or more modalities.

\paragraph{Tabular modality.}
This includes canonical aggregates and bounded summaries such as:
\begin{itemize}
    \item counts and rates,
    \item uniqueness and diversity measures,
    \item rarity indicators,
    \item privilege spread,
    \item role deviation,
    \item local graph statistics,
    \item cross-window deltas.
\end{itemize}

\paragraph{Temporal modality.}
This includes:
\begin{itemize}
    \item short ordered event sequences,
    \item inter-arrival times,
    \item rolling summary histories,
    \item sessionized subsequences,
    \item state-transition traces.
\end{itemize}

\paragraph{Graph modality.}
This includes:
\begin{itemize}
    \item induced typed neighborhoods,
    \item edge labels and directionality,
    \item local motifs,
    \item structural signatures,
    \item provenance-aware edge confidence.
\end{itemize}

\paragraph{Provenance modality.}
For auditability and deployment realism, object construction may also preserve:
\begin{itemize}
    \item source-system identities,
    \item adapter lineage,
    \item confidence scores,
    \item missingness indicators,
    \item partial-observability masks.
\end{itemize}

This modality separation is useful because different downstream learners may consume different subsets while remaining anchored to the same object semantics. It also supports deployment realism: in practice, many systems operate under partial visibility, uneven source quality, or source loss, and CSTS should preserve those conditions explicitly rather than hide them.

\subsection{Views and Semantics-Preserving Transformations}

Contrastive and metric learning require multiple valid presentations of the same underlying object. We therefore define a \emph{view family} over an object $o$ as a collection
\[
\mathcal{V}(o) = \{\widetilde{o}_1,\ldots,\widetilde{o}_m\}
\]
obtained by applying admissible transformations from a view-operator family $\mathfrak{V}$.

\subsubsection{Admissible view operators}

Typical admissible operators include:
\begin{itemize}
    \item \textbf{attribute masking} $\mathcal{M}_{attr}$: hide or perturb selected non-core attributes;
    \item \textbf{neighborhood subsampling} $\mathcal{S}_{nbr}$: sample a bounded subset of neighbors while retaining focal identity and essential relation types;
    \item \textbf{temporal offsetting} $\mathcal{S}_{time}$: construct nearby support intervals around the same focal object;
    \item \textbf{source omission} $\mathcal{S}_{src}$: suppress one telemetry producer or modality while preserving canonicalized structure;
    \item \textbf{modality projection} $\Pi_{tab}, \Pi_{temp}, \Pi_{graph}$: expose only one modality of the same object;
    \item \textbf{confidence-threshold pruning} $\mathcal{P}_{conf}$: remove low-confidence edges or attributes while preserving semantic consistency.
\end{itemize}

\subsubsection{View validity constraints}

A transformation $v \in \mathfrak{V}$ is valid only if:
\begin{enumerate}
    \item focal identity is preserved;
    \item typed relationship semantics are not violated;
    \item temporal support remains coherent;
    \item provenance remains recoverable;
    \item the transformed object retains the same underlying behavioral interpretation.
\end{enumerate}

These constraints are essential. Without them, contrastive or metric learning may encourage invariances that are operationally meaningless or that erase the very threat signal one aims to preserve. The value of contrastive learning in the CSTS setting is precisely that semantically aligned views can be brought close in embedding space while semantically distinct behavior remains separable~\cite{khosla2020supcon,koukoulis2025selfsup,wilkie2026contrastive}.

\subsection{Positive, Negative, and Hard-Negative Policies}

To support contrastive and metric learning over CSTS, it is useful to define pairing policies over learning objects.

\subsubsection{Positive pairs}

A pair $(o_i,o_j)$ is a positive pair when the two objects are distinct valid views of the same underlying behavioral unit or when they instantiate the same semantic role pattern under allowable deployment variation. Common examples include:
\begin{itemize}
    \item two views of the same interaction-state object under different masking patterns;
    \item adjacent-window views of the same entity-state object;
    \item subgraph-state objects representing the same canonical actor under different source visibility;
    \item cross-environment objects occupying the same semantic role under stable CSTS typing.
\end{itemize}

\subsubsection{Negative pairs}

A pair $(o_i,o_j)$ is a negative pair when the two objects represent distinct behavioral units whose relational or temporal semantics should remain separable in embedding space. Examples include:
\begin{itemize}
    \item benign host-state versus attack-associated host-state;
    \item unrelated interaction-state objects with different role semantics;
    \item dissimilar motif-state objects.
\end{itemize}

\subsubsection{Hard negatives}

Hard negatives are especially important in enterprise telemetry. These are object pairs with superficially similar aggregate statistics but materially different relational meaning. Examples include:
\begin{itemize}
    \item two flows with similar volume and duration but different process or credential context;
    \item two hosts with similar degree statistics but different role neighborhoods;
    \item two motifs with similar density but different typed edge composition.
\end{itemize}

CSTS is particularly well suited for hard-negative construction because it makes entity identity, typed relationships, provenance, and temporal support explicit, allowing pairing policies to use semantic context rather than only scalar similarity. This is also where metric-learning formulations remain useful: they help organize latent geometry for low-support, subtle, or fine-grained threat classes~\cite{zdt_metric,martinez2024metricfused}.

\subsection{Contrastive Objectives over CSTS Objects}

Let $f_\theta(o)$ denote an encoder mapping a CSTS object to an embedding vector $z \in \mathbb{R}^d$. Contrastive learning over CSTS objects seeks representations in which semantically aligned views are close and semantically distinct objects are separated.

A generic supervised or self-supervised contrastive objective can be written abstractly as
\[
\mathcal{L}_{con} = \sum_i \ell\left(z_i, \mathcal{P}(i), \mathcal{N}(i)\right),
\]
where $\mathcal{P}(i)$ and $\mathcal{N}(i)$ denote positive and negative sets constructed according to a CSTS-governed pairing policy. The specific loss may be instantiated as supervised contrastive learning, self-supervised contrastive learning, or triplet-style metric objectives~\cite{khosla2020supcon,koukoulis2025selfsup}.

In the CSTS setting, the value of such objectives is not merely generic representation quality. Rather, they encourage embeddings that:
\begin{itemize}
    \item preserve semantic role similarity across environments,
    \item remain stable under partial observability,
    \item separate known malicious manifolds from benign manifolds,
    \item expose geometry useful for open-set and zero-day reasoning.
\end{itemize}

\subsection{Open-Set and Zero-Day Implications}

The zero-day setting motivates a particularly important use of CSTS learning objects. Let $o^{(\mathrm{flow})}$ denote an interaction-state object centered on a focal network flow. Under CSTS, this object can be expanded to include:
\begin{itemize}
    \item source and destination host state,
    \item associated process and credential context,
    \item one-hop or two-hop typed neighborhood,
    \item rolling behavioral summaries over multiple windows,
    \item provenance-aware missingness or visibility masks.
\end{itemize}

A reconstruction-based anomaly detector may score such objects by deviation from learned benign or expected geometry, consistent with prior autoencoder-based anomaly-detection approaches~\cite{sakurada2014anomaly,zdt_flow}. A metric or contrastive learner may further organize embedding space so that known malicious behaviors form structured regions while semantically novel behavior lies at greater distance from both benign and known-malicious manifolds~\cite{zdt_metric,wilkie2026contrastive}. This gives CSTS a direct role in open-set and zero-day architectures: the substrate defines object semantics and view validity, while the downstream learning objective shapes the geometry.

\subsection{Why This Appendix Matters for CSTS}

This appendix clarifies that CSTS is not only a normalization layer but also a \emph{governed representation substrate} for AI-native security learning. Canonical entities, typed relationships, temporal continuity, and provenance are the substrate primitives. Learning objects are bounded constructions over those primitives. Views are semantics-preserving transformations of those objects. Pairing policies and contrastive objectives then operate over that governed space.

This layered interpretation is important for portability. Schema stability and representational invariance are enforced at the substrate layer. Learning-object and view semantics extend those guarantees upward into model construction, reducing the ambiguity under which contrastive, metric, anomaly-detection, and open-set learners are trained and deployed. CSTS does not solve all downstream generalization problems by itself, but it sharply reduces representational ambiguity and provides a principled substrate over which follow-on work in zero-day detection, graph-aware novelty detection, and contrastive cyber representation learning can be built~\cite{zdt_flow,zdt_metric,koukoulis2025selfsup,wu2024egconmix,wilkie2026contrastive}.

%--------------------------------------------------
%
%                   Bibliography
%

\bibliographystyle{IEEEtran}
\bibliography{ref}

\end{document}